# A van der Waals photon-ratchet for enhancing spin-defect emission and quantum sensing

E. Mayner[1*], Y. Zhumagulov[2], C. de Giorgio[3], F. Chu[1], P. P. Swain[4], G.E. Fantner[4], A. Kis[3], O. V. Yazyev[2], A. Radenovic[1*]

[1]LBEN, Bioengineering Institute, École Polytechnique Fédérale de Lausanne (EPFL), CH-1015 Lausanne, Switzerland.

[2]C3MP, Institute of Physics, École Polytechnique Fédérale de Lausanne (EPFL), CH-1015 Lausanne, Switzerland

[3]LANES, Electrical Engineering Institute, École Polytechnique Fédérale de Lausanne (EPFL), CH-1015 Lausanne, Switzerland

[4]LBNI, Bioengineering Institute, École Polytechnique Fédérale de Lausanne (EPFL), CH-1015 Lausanne, Switzerland.

[*]Email: eveline.mayner@epfl.ch, aleksandra.radenovic@epfl.ch

## ABSTRACT

Spin defects in two-dimensional materials are promising platforms for quantum information technologies and nanoscale sensing. The negatively charged boron vacancy ($V_B^-$) in hexagonal boron nitride (hBN) is particularly attractive due to its sensitivity to temperature, magnetic fields, and pressure, but its application is limited by intrinsically weak photoluminescence (PL). Here we demonstrate a van der Waals heterostructure that enhances the PL intensity of $V_B^-$ by up to 45× at 4 Kelvin while preserving compatibility with ultrathin optoelectronic platforms. A type-I band alignment at the $PbI_2$–hBN interface facilitates efficient interlayer coupling to the defect, and strong spectral overlap supports fluorescence resonance energy transfer. Ab initio calculations indicate that this coupling enables a photon-ratcheting mechanism, in which optical absorption in $PbI_2$ generates a nonlinear polarization that drives $V_B^-$ excitation. This process is consistent with interlayer energy transfer while maintaining minimal electronic and spin hybridization and preserving optically detected magnetic resonance (ODMR) contrast. The heterostructure further exhibits improved sensitivity to external magnetic fields. These results establish a route for the enhancement of defect emission via interlayer optical coupling and introduce a general materials strategy for amplifying weak quantum defect signals in nanomaterials.

# INTRODUCTION

Optically addressable spin defects in solids have emerged as powerful platforms for quantum photonic information processing[1–4] and nanoscale sensing.[5,6] Quantum sensing based on these platforms has proven effective across applications and device scales, from detecting individual nuclear spins[7] to measuring currents in full battery systems.[8] Among the most established platforms are color centers in silicon carbide[9,10] and in diamond,[1,11,12] which exhibit robust spin coherence and mature optical control. However, their bulk three-dimensional nature presents intrinsic challenges for integration into planar and scalable device architectures.

Two-dimensional (2D) materials offer a compelling alternative. The optical readout of spin defects in hBN has fundamental and technological benefits for quantum applications:[13,14] Defects can be positioned within nanometers of the surface[15] and this material offers an atomically clean surface with minimal dangling bonds and charge traps.[16] hBN has also long been employed as an encapsulation layer or gate dielectric material in vdW electronics,[16,17] making it naturally compatible with existing device platforms without additional fabrication overhead. More broadly, as quantum sensing expands from controlled dry environments toward biological and liquid-phase settings,[18] the atomically thin and chemically accessible surface of hBN offers advantages for close sensor–sample contact, as demonstrated by optical monitoring of interfacial reactions[19] and quantum detection of paramagnetic species in liquids.[20]

Control of the spin-photon interface has been realized in hBN with several defect candidates, including the negatively charged boron vacancy ($V_B^-$)[21–26] and carbon-related[27,28,7,29] defects. Optical initialization and readout are typically achieved via optically detected magnetic resonance (ODMR), where spin-dependent excitation and relaxation pathways enable the detection of magnetic fields, electric fields, strain, and temperature.[15,22,30] Despite the centers being well-suited for integration into ultra-thin hBN,[31] practical applications remain limited by the relatively low photoluminescence brightness[32,33] and fast decoherence of these defects.[23,34,35] While photonic cavity engineering has demonstrated emission enhancement through Purcell effect,[36–38] such approaches degrade the facile integrability of the defect sensor into flat electronics.

In this work, we demonstrate heterojunction-mediated enhancement of weak spin-defect signals, realizing energy transfer to a spin defect rather than quenching the defect, as reported in related hBN heterostructures.[31] We report that the emission intensity and magnetic-field sensitivity of the $V_B^-$ defect can be enhanced by integrating another layer, lead(II) iodide ($PbI_2$), in a vdW heterostructure. The resulting type-I heterojunction supports short-range interlayer energy-transfer pathways, including dipole-mediated transfer, exchange-mediated transfer and exciton harvesting, alongside nonlinear photon-mediated excitation.

We develop a theoretical framework describing this system as a photon-ratchet, in which a $PbI_2$ layer serves as an excitonic reservoir coupled to the emitting defect in hBN. Low-temperature measurements reveal pronounced enhancements in both absorption and emission, with emission enhancement persisting at room temperature and under low excitation power. The model predicts stronger enhancement ultrathin irradiated-hBN layers, where a larger fraction of defects lies within the short-range interfacial coupling region, while experimentally the largest relative gains are observed in initially dim flakes. These results suggest particular advantages when scaling toward ultrathin and low-signal defect systems. Continuous-wave ODMR measurements further demonstrate improved magnetic field sensitivity across multiple excitation wavelengths.

Together, these results establish vdW interlayer coupling as an effective route for enhancing the optical readout of spin defects. The combined experimental and theoretical evidence supports contributions from nonlinear photon-mediated excitation and interlayer energy-transfer pathways, while preserving the characteristic ODMR response of the $V_B^-$ defect. This materials-based approach provides a scalable fabrication-compatible route for engineering spin–photon interfaces and improving quantum-defect readout in ultrathin sensing platforms.

# RESULTS

## LOW TEMPERATURE MEASUREMENTS AND THEORETICAL BASIS

The results presented in this work were obtained from the exfoliation of a neutron-irradiated bulk crystal of hBN. The $V_B^-$ color centers were generated in the bulk crystal via this neutron irradiation with a dose of $2.6 \times 10^{16}$ n $cm^{-2}$.[23] Because the work concerns the enhancement of the dim PL of the $V_B^-$, to account for flake-to-flake variations in thickness and defect density, heterostructures were prepared with geometries allowing the analysis of isolated materials, with consistent hBN capping between regions. $V_B^-$ emission was then compared with and without the energy donor layer. For clarity, we will refer to normal hBN as hBN and neutron-irradiated hBN as $V_B^-$. In our case, we created four-layer vdW heterostructures (as described in Methods): The top and bottom are hBN capping layers, then the $V_B^-$ layer, followed by a lead(II) iodide layer ($PbI_2$). The heterostructure layers are shown schematically in Figure 1a.

In this system, energetic alignment and spectral overlap enable interlayer coupling. The experimental observations and calculations support a contribution from nonlinear photon-mediated excitation, while short-range pathways including FRET, Dexter exchange, and exciton harvesting may also contribute (Supplementary Section 5.1). $PbI_2$ was selected as the donor material owing to its favourable bandgap (2.4 eV) and band alignment with the $V_B^-$ "bandgap" within the wide bandgap of hBN and its layered crystal structure, which can be mechanically exfoliated and readily integrated into vdW heterostructures.[39,40] $PbI_2$ also emits in the green-blue region of visible spectrum and exhibits high absorption (approaching $10^5$ $cm^{-1}$),[41,42] making it an attractive donor material.

The $PbI_2$ layer is considered the energy donor and sensitizer/antenna, increasing the absorption and emission by and from the acceptor, $V_B^-$. Emission from the $PbI_2$ free exciton (FE) is expected around 510 nm, while broad emission from the $V_B^-$ defect is expected from 750 to 850 nm. Heat maps of the $PbI_2$ FE and $V_B^-$ emission were acquired by spatially resolved spectral mappings with diffraction gratings centred on high-energy and low-energy regions, respectively. The mapped heterostructure is outlined by white dashed lines in the white-light image (Figure 1b). Donor quenching and acceptor enhancement in the overlap region are consistent with interlayer energy transfer: FE emission decreases markedly (Figure 1c), while $V_B^-$ emission increases (Figure 1d). Individual spectra from the scan confirm this trend (Figure 1e): FE intensity drops in the coupled $PbI_2$-$V_B^-$ region (blue) versus capped $PbI_2$ (green), and $V_B^-$ intensity rises in the coupled versus uncoupled $V_B^-$ (red).

The broad peak from 650 to 750 nm observed in the individual $PbI_2$ PL spectrum (and therefore also in the heterostructure spectrum) is referred to as the $PbI_2$ self-trapped exciton (STE) and is thought to arise from inhomogeneities in the crystal lattice. The STE will be discussed later in the text, but its long tails can spectrally overlap with the $V_B^-$ peak. Thus, to avoid overestimating the enhancement of the defect by the heterostructure, it was calculated using differential intensities, as shown in Figure 1f. The background spectrum is subtracted from the $V_B^-$ spectrum, while an adjacent $PbI_2$-only spectrum from the same flake is subtracted from the 4-layer stack spectrum. The $PbI_2$ references were selected to have a comparable STE contribution and to exclude visibly strained regions. The resulting differential defect signal exhibits a 45-fold enhancement at 4 K. Although the differential spectrum obtained from the 4 K map has a limited signal-to-noise ratio, longer-acquisition room-temperature spectra from the same device recovers the characteristic $V_B^-$ lineshape and shows a 6.5-fold photon-count enhancement (SI Fig. 14). Across the measured devices, the largest room-temperature differential enhancement was 15-fold (SI Fig. 4). Residual spatial heterogeneity of the $PbI_2$ emission contributing systematic uncertainty to the extracted value

The low quantum efficiency of the negatively charged boron vacancy is understood by the dipole-forbidden transition between the ground state (GS) and excited state (ES) triplet manifolds in the absence of strain or electric fields.[32] The present work does not seek to modify the intrinsic emission process of the defect, but rather to introduce alternative pathways for optical absorption. This process is described by a photon-ratcheting mechanism, in which optical absorption in $PbI_2$ generates a second-order polarization that drives excitation of the $V_B^-$ defect. In contrast to conventional exciton transfer mechanisms, this process is mediated by nonlinear optical coupling. This optically mediated pathway enhances defect absorption without requiring significant electronic or spin hybridization, thereby preserving the intrinsic radiative and spin properties of the defect.

A representative pathway is illustrated in Figure 2a. $PbI_2$ interacts with the optical field of a high energy photon, $\omega_1$ and emits two photons: a lower-energy infrared photon $\omega_2$, and a photon at the $V_B^-$ defect characteristic transition energy, $\Omega = \omega_1 - \omega_2$. This process arises from the system's second-order nonlinear optical response. The second-order polarization, $P_i^{(2)}$, at frequency, $\Omega$, can be expressed as the product of the second-order susceptibility tensor $\chi^{(2)}$ and the interacting electromagnetic fields, $E(\omega_1)$ laser field and $E(\omega_2)$ infrared photon field:

$$P_i^{(2)}(-\Omega) = \Sigma_{j,k}\chi_{ijk}^{(2)}(-\Omega;\omega_1,-\omega_2)E_j(\omega_1)E_k^*(\omega_2) \quad 1)$$

In the present theoretical treatment, we consider the two-photon nonlinear pathway; phonon-assisted processes and higher-order multiphoton mechanisms are not included. The nonlinear polarization, in turn, governs the $V_B^-$ emission intensity, I, which scales with its magnitude squared.

$$I(\Omega) \propto \left|\mathbf{P}^{(2)}(\Omega)\right|^2 \quad 2)$$

The details of first-principles calculations are provided in the Supplementary Information Section 2. The calculated spin density distribution at the defect in the coupled heterostructure is shown in Figure 2b. The calculated spin density at the $V_B^-$ defect in isolated hBN and hBN-$PbI_2$ heterostructure are shown in SI Fig. 1. The result for the isolated hBN is consistent with previous theoretical studies.[21] The spin-resolved band structures are also shown in SI Fig. 2. These results indicate that the electronic and spin structure of the $V_B^-$ defect remains largely preserved upon coupling to $PbI_2$, consistent with the experimentally retained ODMR resonance positions and preservation of spin-dependent optical readout in the coupled structure. Band structure and hybridized structure will be discussed later in the text. Simplified modelling based on a 7×7 spin Hamiltonian is presented in Supplementary Section 6, wherein changes in eigenstate composition and basis-state mixing are calculated before and after $PbI_2$ coupling.

Increased absorption by the heterostructure for energies below the $PbI_2$ band gap was also observed. In the differential reflectance scan in Figure 2c, which was performed for 550-555 nm at 4K, the heterostructure region shows a lower differential reflectance (blue) than either the $PbI_2$ or $V_B^-$ layers alone. In Figure 2d, the theoretical absorption by the defect, when uncoupled, is compared with the defect absorption, when coupled with $PbI_2$. There is a pronounced boost in absorption of energies higher than the $PbI_2$ bandgap. However, these enhancements also extend to energies below the $PbI_2$ bandgap, indicating the presence of nonlinear optical coupling. Emission enhancement was also calculated from first principles. The relation between excitation energy and PL is nontrivial, showing hyperbolic peaks and troughs (Figure 2e). Outside of the troughs, the enhancement (the ratio of the predicted $V_B^-$ PL with and without the $PbI_2$ layer at the same excitation wavelength) is estimated to be around one order of magnitude, consistent with the experimentally observed enhancement, ranging from 4-15× (room temperature) to 45× (4 K). Photoluminescent excitation (PLE) data also shows increased defect emission (emission collected after a 775 nm LP filter) when exciting with energies lower than the $PbI_2$ bandgap (SI Fig. 3).

## ROOM TEMPERATURE PHOTOLUMINESCENCE

After confirming enhanced absorption and photoluminescence in the coupled structure both theoretically and at low temperature, we assessed performance at room temperature (RT) to evaluate suitability for practical sensing under ambient conditions. The heterostructure exhibits the same characteristic donor quenching and acceptor enhancement at RT (Figure 3a–c) as observed at 4 K (Figure 1b–d). Overall, we observed an enhancement range of 4 to 15× at room temperature (SI Fig. 4), determined in the same manner as described for Figure 1. The differential spectra are shown in SI Figure 4 for the room temperature PL shown in Figure 3. The confocal measurements of PL spectra are described in Methods.

In Figure 3c, high intensity is also observed in $PbI_2$-only regions. These regions coincide with strained areas of the $PbI_2$ flake visible in the optical image in Figure 3a. Spectra extracted from these locations reveal a red-shifted $PbI_2$-associated broadband component extending into the 780–820 nm window (SI Fig. 10). Possible contributions from local lattice distortion, stacking, and defect-related states are discussed in the Extended Materials Characterization section of the SI.

While the size and vacuum-level position of the $PbI_2$ bandgap are relatively well established, the $V_B^-$ "bandgap" was originally inferred theoretically from previous studies (forming the basis of our choice of $PbI_2$ as a quasi type I aligned semiconductor),[43,44,21] the electronic bands of the two materials obtained from DFT calculations are presented in Figure 3d. The $PbI_2$ valence band is well below the highest occupied state of the defect, such that the electron will travel to a lower energy state when migrating from $PbI_2$ to the $V_B^-$ state in hBN. On the other hand, the conduction band states are close in energy for the two materials. Based on the known errors of DFT energy calculations and the experimental results indicating ET, the band alignment is understood to promote energetically favorable transfer of excitation from $PbI_2$ to $V_B^-$, type I band alignment. A simplified diagram for this process is shown in Figure 3f, wherein the energy from vacuum for the valence and conduction bands of each material are labelled. The bandgaps are placed around the Fermi levels, determined by Kelvin Probe Force Microscopy (KPFM) in Figure 3e, performed in vacuum overlayed on its SEM image. While reverse transfer into the $PbI_2$ free-exciton or band-edge state is energetically unfavorable, sufficiently red-shifted localized $PbI_2$ states could in principle provide lower-energy acceptor states; such reverse transfer would act as a competing loss channel rather than explain the observed $V_B^-$ enhancement.

KPFM probes the Fermi level relative to vacuum. The area off of the structure is gold, and its Fermi level is 5.1 eV relative to vacuum. The KPFM measurement procedure and calibration are detailed in Methods. To determine the effect of band-bending during stacking, the structure fabricated for KPFM has areas where one, two, and three layers exist. The results indicate, as predicted from this work and previous theoretical works,[43,21] that the Fermi levels of the materials are well aligned (all within 0.5 eV) to promote energetically favorable transfer of excitation. FRET is also favorable in the coupled structure, supported by the experimental emission and absorption spectrum in Figure 3g. Absorption spectra were recorded using a bulk neutron-irradiated hBN crystal.

The defect lifetime was then measured in both the coupled and uncoupled regions of the heterostructure using a 775 nm long-pass (LP) filter to suppress $PbI_2$ emission. At 4 K under 460 nm excitation, the coupled region exhibits a longer lifetime than the irradiated-hBN reference, whereas at room temperature under 555 nm excitation, it exhibits a shorter lifetime (SI Fig. 6a,c). The room-temperature difference may reflect a greater contribution from $PbI_2$-associated broadband emission, while the longer low-temperature lifetime may indicate delayed feeding of $V_B^-$ from the $PbI_2$ exciton reservoir; however, residual spectral contributions cannot be excluded. The decays were fitted with IRF-deconvoluted biexponential functions, and the intensity-weighted average lifetime was used for comparison.

To investigate the recombination mechanisms within the heterostructure, power-dependent PL measurements were performed from 15 to 160 μW. Excitation and acquisition parameters were chosen to maintain defect signal visibility while avoiding saturation of the $PbI_2$ free-exciton emission. In the $PbI_2$ region, two primary features are observed (Figure 4a): the free exciton and the broad emission feature spanning 600 to 750 nm, assigned to the self-trapped exciton. STE emission arises from exciton localization through crystal lattice distortion,[45] a phenomenon that is also linked to iodine vacancies.[46] This localization significantly redshifts the emission and results in a broad spectral feature due to multiple vibronic transitions.

Next, the structural inhomogeneities of $PbI_2$ were evaluated using high-resolution TEM and elemental analysis (Figures 4b and 4c, respectively). The TEM reveals regions of varying thickness and local lead accumulation, indicating substantial structural and compositional heterogeneity that may contribute to the observed broadband emission. We then examined how the different emission components evolve with increasing excitation power. Figure 4d presents power-dependent emission spectra from the $PbI_2$-coupled region of the structure, showing a continuous increase in FE, STE, and $V_B^-$ emission with increasing excitation fluence. The differential intensity (het-$PbI_2$, as in Figure 1f) as a function of laser power is also shown in the inset. The absence of roll-off from the $V_B^-$ emission indicates that the defect ensemble remains unsaturated within the power range. The corresponding power-dependent PL from the irradiated-hBN reference region is shown in SI Fig. 8. Power-law analysis (Figure 4e) yields similar scaling exponents for coupled and uncoupled regions, indicating that $PbI_2$ does not alter the intrinsic radiative efficiency of the defect ensemble, consistent with our section on the ET mechanism. Peak intensities were extracted using two-Gaussian fits for heterostructure spectra and single-Gaussian fits for neutron-irradiated regions (SI Fig. 8).

To understand how this behaviour persists with different initial defect brightness, Figure 4f quantifies ensemble-level coupling by plotting the change in $V_B^-$ emission intensity against baseline defect PL. The observed linear scaling shows that the absolute enhancement increases with the intrinsic defect signal, consistent with a larger number of sensitized defects rather than saturation of the defect ensemble. Given the effectively unlimited $PbI_2$ exciton reservoir and the persistence of the $PbI_2$ FE peak observed in the coupled structure, this behaviour supports a transfer-limited regime, in which a substantial fraction of donor excitons recombine through competing pathways.

Finally, we examined how the donor efficiency influences the coupled emission as a function of temperature. Temperature-dependent PL (Figure 4g) measurements reveal a multi-order-of-magnitude increase in $PbI_2$ FE emission upon cooling, accompanied by a corresponding enhancement of $V_B^-$ defect luminescence in the heterostructure. This correlation indicates that improved donor-exciton efficiency strengthens defect sensitization and is consistent with the larger enhancement observed at 4 K than at room temperature.

## Sensing Magnetic Fields with the Enhanced Emission

The $V_B^-$ defect is primarily of interest as a qubit platform, here as a sensor. The $V_B^-$ electron spin can be off-resonantly excited from its spin-triplet ground state to its spin-triplet excited state via optical absorption/laser irradiation. Because the non-radiative, spin flip decay channel (accessed via intersystem crossing) preferentially couples to the $m_s = \pm1$ sublevels of the excited states, excitation of the $m_s = \pm1$ states results in lower PL. A simplified energy diagram for the $V_B^-$ qubit is shown in Figure 5a, where $D_{gs}$ is the room-temperature natural ground-state splitting between sublevels $|0\rangle$ and $|\pm1\rangle$ and $D_{es}$ is the room-temperature natural excited-state splitting between sublevels $|0\rangle$ and $|\pm1\rangle$. This difference in PL enables one to read out the electron spin state and is known as the contrast.

Optically detected magnetic resonance (ODMR) is used here to probe these states. In ODMR, the PL is monitored while sweeping the frequency of the microwave delivered to the sample via antenna. The setup is illustrated schematically in Figure 5b, and ODMR methods are further described in Methods. When the frequency crosses a transition where the energy is equivalent to the energy between the ground state electron spin sublevels, the resonance is evidenced as a drop in the PL signal. The full form of the resonant features, including an external magnetic field, $B$, is described by Equation 3:

$$v_{1,2} = \frac{D_{gs}}{h} \pm \frac{1}{h} * \sqrt{E_{gs}^2 + (g\mu_B B)^2} \qquad 3)$$

Here $D_{gs}$ and $E_{gs}$ denote the ground state longitudinal and transverse zero-field splitting parameters, respectively. $h$ is Planck's constant. $\mu_B$ is the Bohr magneton, and $g$ is the Landé factor (2). The PL counts were recorded as a function of microwave frequency, comparing different regions of the heterostructure. The system was excited off-resonantly by either a 488 nm (Figure 5c) or 555 nm (Figure 5d) laser, and, in both cases, we observe a strong improvement of the PL signal in the heterostructure region while retaining the same resonant features as in the $V_B^-$ region.

To determine the sensing capability of the defect, we applied an external magnetic field and measured the corresponding shift in the resonant ODMR peaks. Selected datapoints are plotted on a normalized scale in Figure 5e, showing the expected characteristics: As the magnetic field increases, the separation of the resonance peaks increases. Peaks are shown with their full contrast in SI Fig. 11. Magnetic fields vs the resonance frequencies were then plotted with additional points as a heat map (Figure 5f), demonstrating the ability to sense the field using the resonance position of ODMR peaks. The theoretical resonance frequencies (from Equation 3) vs the magnetic field are plotted in light blue and clearly agree with the experimental results. From this work, we derive $D_{gs}/h$ = 3.46 GHz and $E_{gs}/h$= 55 ±5 MHz, which is in good agreement with previous works on the $V_B^-$.[23,47]

To quantify the impact of the $PbI_2$ ratcheting layer on $V_B^-$ utility, we evaluated the continuous-wave ODMR (cw-ODMR) sensitivity, which is defined in Equation 4.

$$\eta_{CW} = \frac{4}{3\sqrt{3}} \frac{h}{g_e \mu_B} \frac{\Delta v}{C_{CW}\sqrt{R}} \qquad 4)$$

Sensitivity ($\eta_{CW}$) depends on the ODMR linewidth ($\Delta v$), contrast ($C_{CW}$), and photoluminescence count rate ($R$). All measurements were performed at room temperature; to minimize the impact of $PbI_2$ STE emission, a 775 nm LP filter was used. Because a substantial fraction of $V_B^-$ emission lies below this cutoff, the reported enhancement represents a conservative lower bound. Under 488 nm excitation, the heterostructure exhibits an increase in PL counts while preserving resonance positions (Figure 5c); under 555 nm excitation (below the $PbI_2$ bandgap), a threefold increase is observed (Figure 5d). This photon-collection gain, together with moderate changes in contrast, results in a twofold sensitivity improvement at 488 nm and greater than threefold at 555 nm (SI Tables 3–4). The heterostructure reaches the same sensitivity as the irradiated-hBN reference region using less than one-third of the excitation power (SI Fig. 12). It also remains consistently more sensitive than the reference region over ten days, following long-term glovebox storage and repeated ambient exposure between measurements (SI Fig. 13).

The best-performing (highest sensitivity) heterostructure in this work achieved a sensitivity of 88 μT $Hz^{-1/2}$ at 160 μW (SI Table 3), approximately threefold better than the previously reported optimized $V_B^-$ sensitivity measured at the lowest reported excitation power, which nevertheless exceeded the power used here.[48] The absolute sensitivity remains below that of established NV-diamond magnetic microscopes, which can reach single-digit μT $Hz^{-1/2}$ sensitivity at micrometer spatial resolution.[49] Nevertheless, hBN-based sensors offer a complementary advantage through their atomically thin, planar geometry and compatibility with vdW heterostructures and planar device architectures. Whereas shallow or scanning NV geometries often involve sensor–target distances on the order of ~10 nm,[8] hBN may, in principle, reduce this separation toward the vdW scale. The enhancement demonstrated here therefore represents a step toward practical two-dimensional quantum sensors that can be incorporated directly into heterostructures and planar devices.

A larger relative sensitivity enhancement (with 488 nm excitation) was observed for the heterostructure shown in Figure 1, whose irradiated-hBN reference region exhibited a relatively low initial photon count rate of $8.45\times10^4$ $s^{-1}$ and a baseline sensitivity of 368 μT $Hz^{-1/2}$. During ODMR measurements, this device exhibited a 6.5-fold photon-count enhancement and a corresponding 2.6-fold sensitivity improvement (SI Table 5 and SI Fig. 14). The close agreement between the measured sensitivity improvement and the expected square-root scaling with photon count indicates that the additional photons were converted efficiently into sensing performance.

More generally, the room-temperature PL enhancement is lower than at 4 K, consistent with reduced radiative efficiency at elevated temperature. The sensitivity improvement also varies among devices, likely reflecting device-to-device differences in defect density and layer thickness, for which representative AFM measurements are shown in SI Fig. 9, as well as variations in $PbI_2$-associated broadband emission that can red-shift into the detection window (SI Section 4.2). These spin-independent photons increase the detected count rate but dilute the ODMR contrast, such that the largest photon-count enhancement does not necessarily yield the greatest sensitivity improvement. The results therefore identify control of the $PbI_2$ spectral response and thickness, together with optimized filtering, as practical routes for improving room-temperature performance (SI Section 5.1). Performing the entire exfoliation and transfer process under inert conditions may further limit $PbI_2$ degradation and suppress broadband red and near-infrared emission.

Although the present devices span a limited thickness range and do not establish a unique optimum, the steep exponential or inverse-power-law distance dependence of the relevant interlayer interactions identifies a clear route for further improvement. The toy model shown in SI Fig. 16 predicts that the enhancement should be strongest for ultrathin neutron-irradiated hBN flakes with $t_{hBN} \lesssim 10$ nm, where a larger fraction of defects lies within the short-range coupling region near the $PbI_2$ interface. Experimentally, the largest gains are also observed in initially dim irradiated-hBN flakes, suggesting that the heterostructure is particularly effective where the intrinsic defect signal is weak. Reducing the irradiated-hBN thickness should increase the fraction of defects strongly coupled to $PbI_2$ and may therefore produce disproportionately larger gains in ultrathin structures. Combined with improved control of $PbI_2$ thickness and spectral quality, as well as optimized filtering of its broadband emission, this provides a practical strategy for further increasing both photon collection and ODMR sensitivity. Systematically engineered heterostructures could therefore extend the present approach to weak signals from ultrathin or low-defect-density sensing layers.

Overall, these measurements and calculations demonstrate that directed vdW heterostructure engineering can provide a materials platform that enables optically mediated excitation of spin defects through nonlinear photon-mediated excitation and interlayer energy transfer from $PbI_2$ to $V_B^-$ defects in hBN, with possible contributions from other short-range coupling channels. This approach yields substantial enhancement of both photoluminescence and sensing performance of the $V_B^-$ defect, with a 45× increase in emission at 4 K and 4-15× increase persisting at room temperature. Reciprocal photoluminescence mapping reveals the suppression of $PbI_2$ free-exciton emission concurrent with amplification of $V_B^-$ luminescence, supporting interlayer energy transfer. Excitation-dependent PL and absorption measurements further support a contribution from nonlinear optical coupling, in agreement with the density functional theory calculations. Continuous-wave ODMR measurements demonstrate significant gains in magnetic sensitivity across multiple excitation wavelengths at room temperature, a practical step towards a viable 2D quantum sensor. Together, these results establish a generalizable materials strategy for optical enhancement of quantum defects and engineering spin–photon interfaces in ultrathin quantum sensing platforms.

# METHODS

## Heterostructure Fabrication

Commercially available hBN bulk crystals (HQ Graphene) were neutron irradiated at facilities at OSU to generate $V_B^-$ defects. A dose of $2.6 \times 10^{16}$ n $cm^{-2}$ was used.[23] $PbI_2$ bulk crystals were also purchased from HQ Graphene and kept under inert conditions in an argon glovebox. They were purchased in November 2024 and the work took place in the subsequent 1.5 years. The heterostructure was fabricated by a dry-transfer technique using polycarbonate (PC) membranes. Individual flakes were mechanically exfoliated onto $SiO_2$/Si substrate with oxide thickness of 270 using the cleavage technique method pioneered for the production of graphene.[50] The same mechanical exfoliation procedure was used for $PbI_2$ and native hBN. However, because $PbI_2$ flakes were difficult to pick up from $SiO_2$/Si substrates, $PbI_2$ was exfoliated onto Gel-Pak. The heterostructures were assembled by sequentially picking up the individual flakes with a polycarbonate (PC) stamp. During exfoliation and transfer, the $PbI_2$ flakes were exposed to ambient conditions for up to approximately 6 h. The completed stack was released onto the ODMR antenna by heating to 200 °C, causing the PC to pass through its glass-transition temperature and remain on the target substrate.[51] The residual PC was subsequently removed using chloroform, followed by isopropanol. Gel-Pak and $SiO_2$/Si substrates containing exfoliated flakes were stored in an argon glovebox between fabrication steps. Encapsulation with hBN reduced subsequent exposure of the $PbI_2$ to ambient conditions, as supported by the capped–uncapped comparison in SI Fig. 7 and the stable ODMR sensitivity measured over ten days in SI Fig. 13.

## Confocal Microscopy

The PL properties were analysed with a home-built scanning confocal microscope operating at ambient conditions. A laser excitation at 405 nm, 488 nm, or 555 nm (Omicron BrixXHub-6) was focused onto the sample with a high numerical aperture microscope objective (NA = 0.95) mounted on a three-axis MCL Nanodrive piezoelectric scanner (MCLC02485) and an MCL Microdrive (MCL-microD2035). A 4f imaging configuration was used before directing the excitation to the sample via the appropriate long-pass dichroic mirror. The PL signal was collected by the same objective, filtered via a long pass filter with wavelength depending on the application, and focused onto a 50-µm-diameter pinhole, and finally directed either to a spectrometer (Ocean Optics QEPro High Performance Spectrometer) or to a silicon avalanche photodiode operating in the single-photon counting regime (PerkinElmer Optoelectronics Photon Counting Module, Model: SPCM-AQR-14-FC). Single-point spectra, ODMR, or x-y PL scans were performed. The lateral spatial resolution of the microscopy was around 1 µm. For differential PL analysis, reference spectra were selected from spatially adjacent regions within the same spectral map under identical acquisition conditions. $PbI_2$-only regions with comparable STE amplitudes were used, while visibly strained regions with strongly red-shifted broadband emission were excluded.

## Optically Detected Magnetic Resonance

Optically detected magnetic resonance (ODMR) spectra were recorded by monitoring the PL signal while sweeping the frequency of a microwave field applied through a gold coplanar waveguide (antenna) onto which the heterostructure was transferred. The gold antenna was patterned on the coverslip using photolithography and electron beam evaporation of 5 nm titanium and 50 nm gold. The antenna was then wire bonded to a copper coplanar waveguide printed circuit board (PCB) that interfaces with the signal generator (Siglent SSG5060X-V) interfaced with a copper waveguide PCB to the sample surface. When the microwave frequency crosses a transition where the energy is equivalent to the energy between the ground state electron spin sublevels, the magnetic resonance is evidenced as a drop of the PL signal. At zero external magnetic field, the resonance frequencies are given by $\nu_{\pm} = D \pm E$, where D and E denote the longitudinal and transverse zero-field splitting parameters, respectively. The external magnetic field was achieved with a NdFeB magnet (N35) that was positioned above the sample stage and could be raised and lowered with a Z-axis positioner.

## Kelvin Probe Force Microscopy

Kelvin probe force microscopy[52] (KPFM) measurements on the 2D heterostructures were performed using a tip-scanning atomic-force microscope (AFM) inside a scanning electron microscope (SEM).[53] Self-sensing, piezoresistive tri-layer cantilevers with an integrated silicon tip were used to eliminate optical laser beam-induced effects on surface potential and enable measurements in a compact in-vacuum environment. The tip side was coated with Ti/Pt, resulting in a ~100 nm thick Pt layer.[54] The cantilever deflection was read out via the integrated piezo-resistive bridge circuits and used for topography feedback in amplitude-modulation AFM mode. KPFM was implemented in a heterodyne configuration:[55] an AC bias $V_{ac}$ was applied between tip and sample at a frequency $f_{el}$ chosen such that the resulting electrostatic force detection is at the second eigen mode $f_2 = f_1 + f_{el}$, $f_1$ being the first eigen mode of the cantilever and the mechanical carrier frequency. The KPFM feedback loop nulled the electrostatic interaction by adjusting the DC bias $V_{dc}$ until the electrostatic response was minimized, and the resulting compensating bias was recorded as the contact potential difference (CPD) map alongside topography in a single pass.

Sample navigation combined a bright-field reference image, a brief low-magnification SEM survey, and the AFM coarse-positioning stage. Before insertion into the correlative AFM/SEM setup, the sample was mounted with approximately the same orientation as in the optical image, and the AFM tip was positioned near the expected heterostructure-containing region. After chamber pump-down, an SEM survey with a sub-mm field of view was used to locate a nearby group of heterostructures and to register the sample position approximately against the optical reference. SEM scanning was then stopped, and the AFM coarse-positioning stage was used to move the tip to the selected target based on its relative position. The target heterostructure was outside the SEM survey raster and was not imaged with SEM before completion of the KPFM measurements. To reference the work function, the 2D material flakes were prepared on an evaporated Au film that served as an internal reference region within each scan. The large-area CPD map was reconstructed by acquiring a tiled dataset of 9 partially overlapping frames of 20 µm x 20 µm. Before each frame, the bias-dependent CPD response was checked to confirm stable potential nulling. Contact-potential stability was assessed using the Au regions present in each scan. Before Au referencing, the recorded Au signal showed a peak-to-peak variation of 265 mV across the complete tiled acquisition. In one frame containing Au scanned over ~ 80 min of the AFM's slow-scan direction, the Au signal showed a peak-to-peak variation of 78 mV, which was treated as an upper-bound estimate of the intra-scan drift, a combination of instrumental drift, tip-related changes, spatial variation of the Au surface, or contamination of the surface. Each frame was referenced independently to clean Au measured within the same frame before stitching. The frames were registered by translational alignment of common features in the overlapping regions and combined into a single mosaic. A line-dependent Au correction applied separately to all nine complete frames changed the reported region-averaged values by only 4-15 meV and did not affect the values at the stated precision of 0.1 eV. Fermi level values were calculated from CPD values using $\mathrm{CPD} = (\phi_{tip} - \phi_{sample})/e$, where ϕ is the work function and e is the elementary charge. With the tip work function calibrated against Au as reference in each scan, we computed the referenced potential offset as $\Delta V(x,y) = V_{CPD}(x,y) - \langle V_{CPD} \rangle_{Au}$, where $\langle V_{CPD} \rangle_{\mathrm{Au}}$ is the mean CPD measured for Au. The local sample work function was then obtained as $\phi_{sample}(x,y) = \phi_{Au} - \Delta V(x,y)$, assuming $\phi_{Au} = 5.1$ and the equivalence $1V = 1\ eV/e$. The Fermi levels relative to the vacuum level were reported as $E_F(x,y) = -\phi_{sample}(x,y)$.

# Data Availability

The data generated in this study have been deposited in the Zenodo database 10.5281/zenodo.22796827.

## ACKNOWLEDGMENTS

E.M. and A.R. would like to thank Dr Arianna Marchioro (LBP and EPFL) for the photochemical mechanism discussions and experimental advice and Dr David Reyes (CIME EPFL) for TEM imaging, analysis, and overall expertise on sample prep for difficult samples such as $PbI_2$. E.M and A.R. would also like to thank CMI EPFL for the technical support and facilities. Several scientific colour map (vik, cork, batlow) were used in this study to prevent visual distortion of the data and exclusion of readers with colour-vision deficiencies.[56]

## FUNDING STATEMENT

E.M and A.R. acknowledge support by the European Research Council (grant 101020445-2D-LIQUID). Y.Z. and O. V. Y. acknowledge support by the Swiss National Science Foundation (grant 224624). First-principles calculations have been performed at the Swiss National Supercomputing Centre (CSCS) under Project No. lp96 and the facilities of the Scientific IT and Application Support Center of EPFL. P.P.S and G.E.F acknowledge the support from the Technology Agency of the Czech Republic (TACR FW10010168).

## AUTHOR CONTRIBUTIONS STATEMENT

E.M. and A.R. conceived the idea for heterostructure-enhanced defect emission. E.M. performed the experimental work and coordinated the study. Y.Z. carried out the DFT calculations and most of the theoretical modelling, in discussion with O.V.Y. F.C. fabricated the microwave antennas in the cleanroom under the supervision of E.M. Low-temperature (4 K) measurements were led by C.d.G., with guidance from A.K. KPFM measurements were led by P.P.S., with input from G.E.F.

## COMPETING INTERESTS STATEMENT

The authors declare no competing interests.

# FIGURE LEGENDS/CAPTIONS

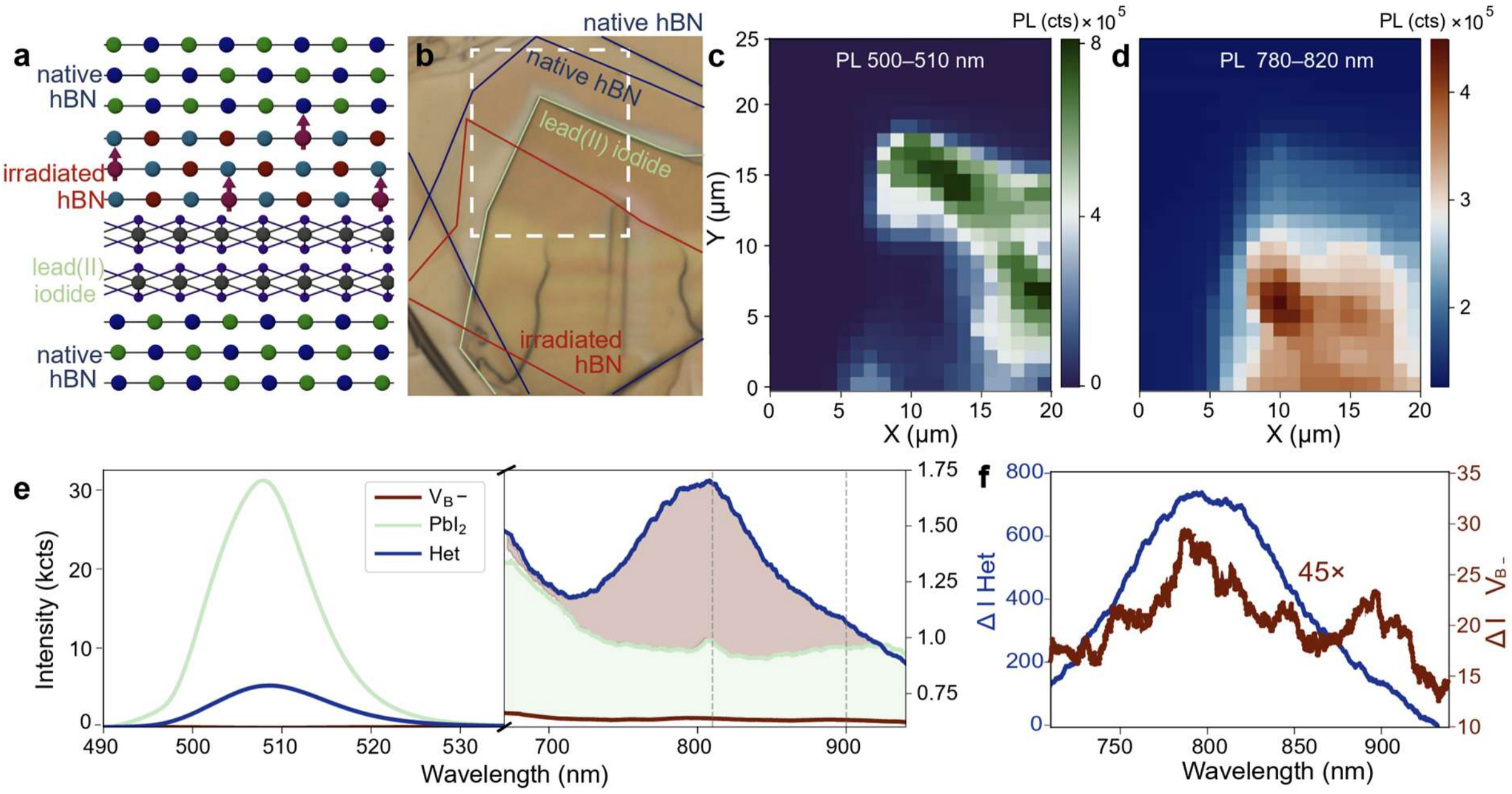

**Figure 1 Enhanced defect emission in hBN–PbI$_2$ heterostructures at 4K: a)** Schematic of the four-layer heterostructure with layers labeled on the left. Boron and nitrogen atoms in neutron-irradiated and native hBN are depicted in distinct colours for clarity. **b)** White-light image of a heterostructure. The top and bottom layers consist of native hBN, serving as capping layers and marked in dark blue. Immediately beneath the top hBN is the neutron-irradiated hBN layer, containing negatively charged boron vacancies, marked in red. This is followed by PbI$_2$, outlined in light green. The area indicated by white dashed lines corresponds to the region that was mapped at a temperature of 4 Kelvin. **c)** The PL intensity map displays the signal associated with the FE peak of PbI$_2$ for a wavelength range 505 ± 5 nm when excited with a 460 nm laser. **d)** The PL intensity map for the $V_B^-$ defect peak was measured at a range of 800 ± 20 nm. **e)** PL spectra corresponding to the regions identified in panels c and d are plotted. PbI$_2$ displays a pronounced FE peak and a STE peak. **f)** The off-flake background was subtracted from the irradiated-hBN spectrum, while adjacent PbI$_2$-only spectrum was subtracted from the full-stack spectrum. The enhancement factor was calculated as the ratio of the peak amplitudes, yielding 45× at 4 K. Both spectra were extracted from the same spectral map using identical acquisition conditions.

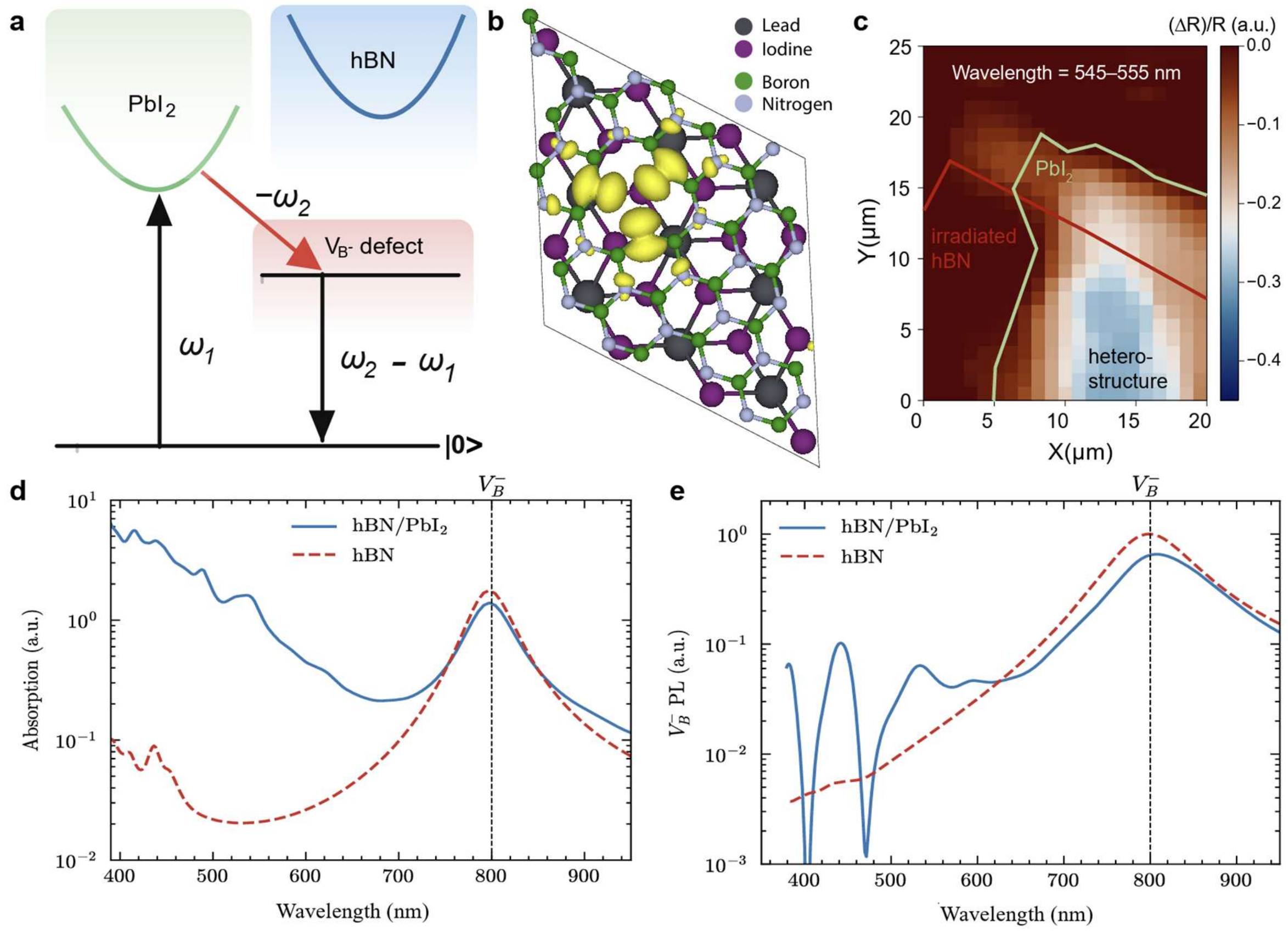

**Figure 2 Mechanism and predicted optical response of $V_B^-$ defects. a)** Schematic drawing of the photon-ratcheting mechanism for emission enhancement, showing nonlinear optically mediated excitation of the $V_B^-$ defect via $PbI_2$. Here, $|0\rangle$ denotes the ground state of the $V_B^-$ optical transition and is distinct from the $m_s = 0$ spin sublevel used in the ODMR description. **b)** Spin density distribution at the $V_B^-$ defect in the hBN-$PbI_2$ coupled heterostructure. The atom labels are in the legend. **c)** Differential reflectance (absorption) observed at 4K from the heterostructure in Figure 1. The coupling between $PbI_2$ (green outline) and $V_B^-$ (red outline) increases absorption. **d)** Expected absorption of the defect in an isolated hBN layer (red dashed line) and coupled with lead iodide (blue line) calculated from first principles. Absorption increases in the coupled structure for photon energies above the $V_B^-$ emission energy, corresponding to excitation wavelengths shorter than the defect-emission wavelength, which is demarcated with a vertical dashed black line. **e)** Theoretical PL intensity of the uncoupled defect (red dashed line) and $PbI_2$-$V_B^-$ coupled structure (blue line) as a function of excitation wavelength. The emission energy line of the defect is indicated with a vertical dashed black line.

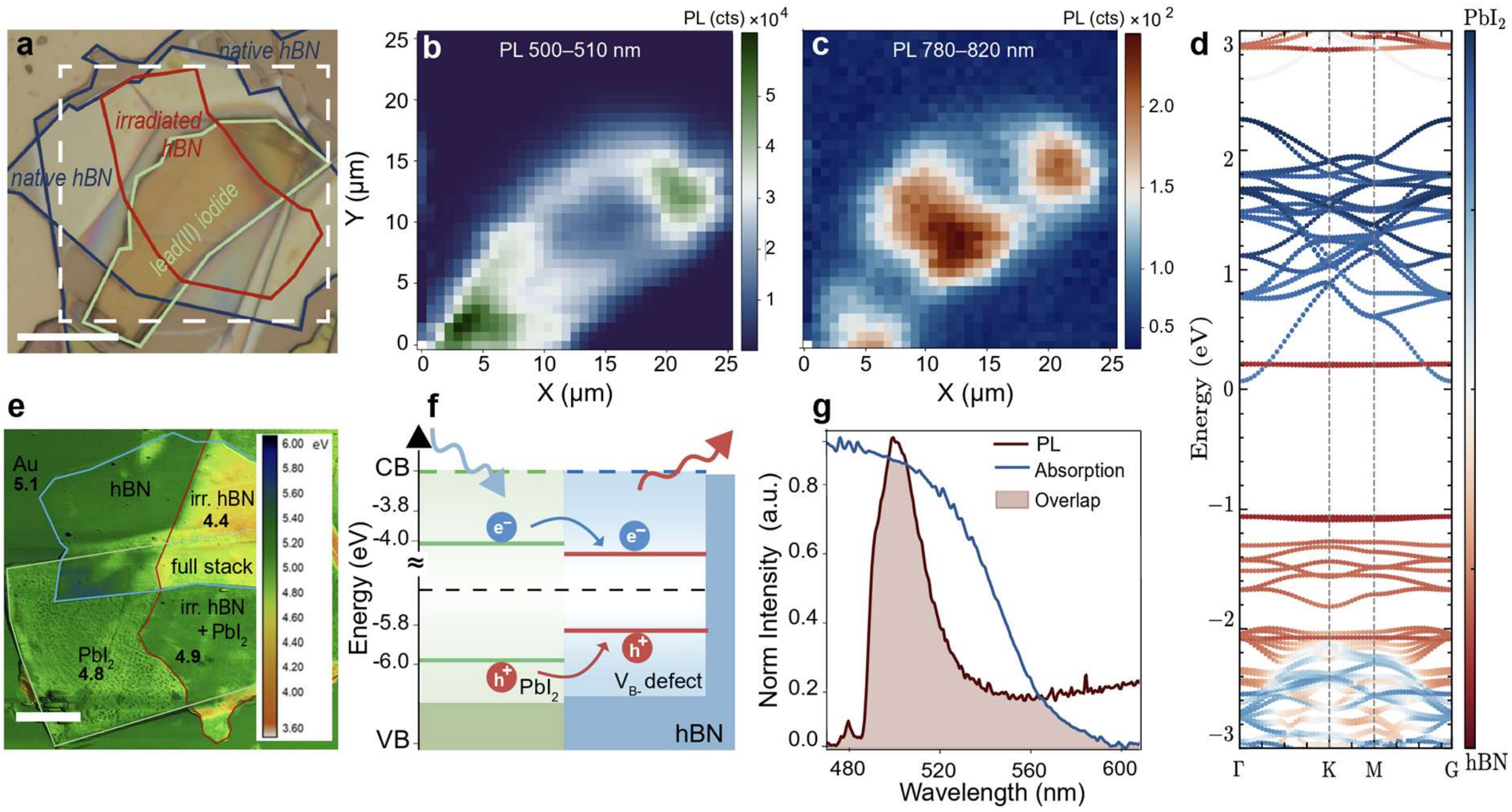

**Figure 3 Room-temperature heterostructure PL characteristics: a)** A white-light image of the imaged four-layer vdW heterostructure. The top and bottom layers consist of native hBN flakes serving as capping layers and are marked in dark blue. Beneath the top hBN is the neutron-irradiated hBN layer, marked in red. This is followed by $PbI_2$, outlined in light green. The area in white dashed lines was mapped at RT. The scale bar is 5 µm. **b)** The PL intensity map of the signal associated with the free exciton peak of $PbI_2$, 510 ± 10 nm. The sample was excited with a 488 nm laser using 48 µW. **c)** The PL intensity map for the $V_B^-$, detected at 800 ± 20 nm. **d)** DFT calculated band structures for $PbI_2$ (blue) and $V_B^-$ in hBN (red), where white refers to the strongly hybridized states. **e)** SEM image overlaid with KPFM data of overlapping layers of $PbI_2$ (bottom, red), $V_B^-$ (purple, middle), and hBN (blue, top). **f)** Simplified scheme of quasi type I (straddling) band alignment, where the band gaps sizes are based on calculations and centered around the Fermi levels determined by KPFM. **g)** Emission and absorption overlap integral required for FRET. Emission of $PbI_2$ is in red and absorption of neutron-irradiated hBN is in blue.

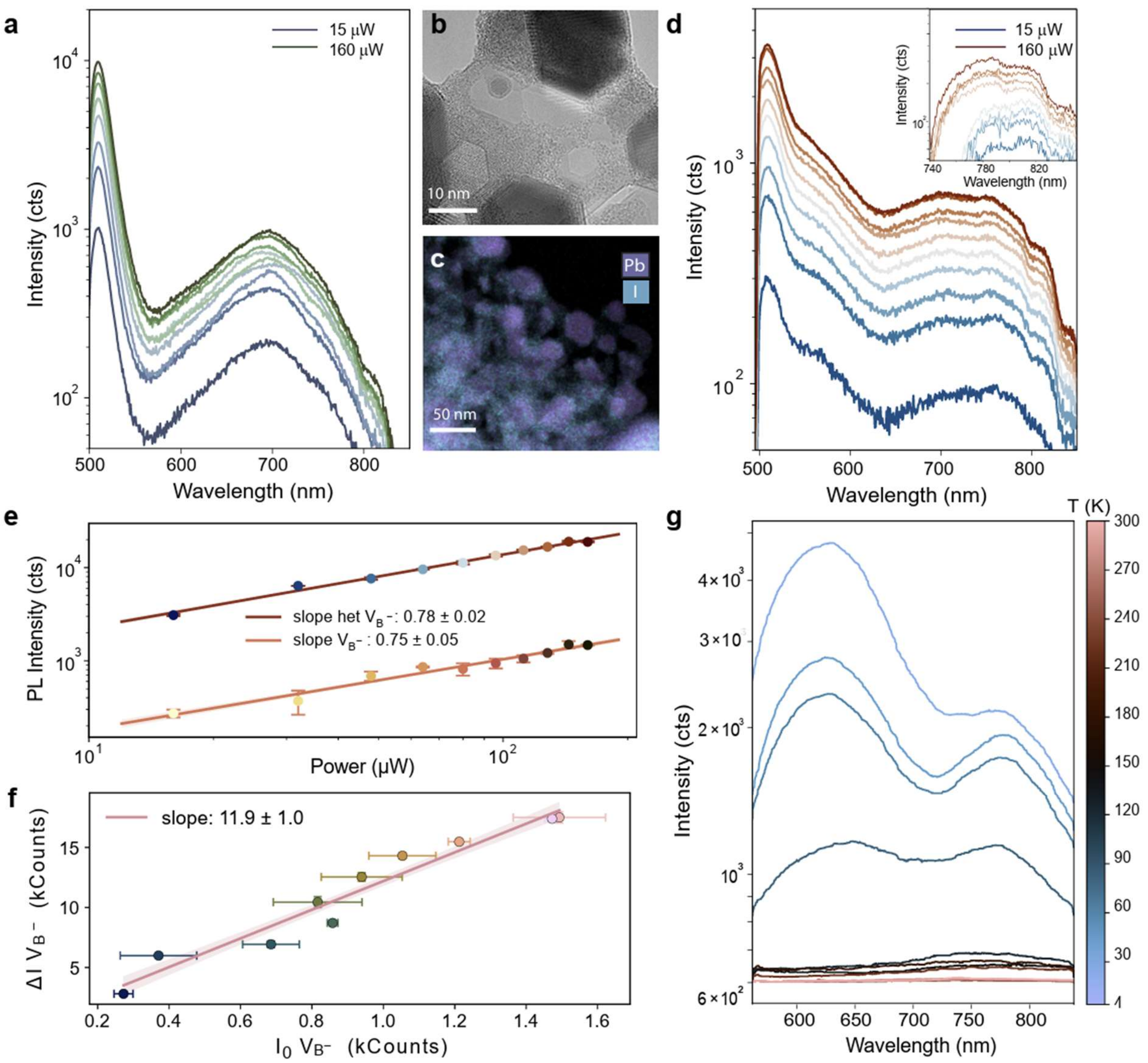

**Figure 4 Emission Characteristics: a)** PL versus power for the $PbI_2$ (capped with hBN) region of heterostructure, referred to as $PbI_2$. The background (off-heterostructure) was subtracted. Excitation was at 488 nm, with power increasing from 15 µW to 160 µW. The endpoint colors corresponding to 15 and 160 µW are indicated in the legend, and intermediate traces correspond to successive 10% increments of the applied laser power (power density 1.9–20.4 kW $cm^{-2}$). **b)** High-resolution TEM of a $PbI_2$ flake is shown, demonstrating local structural inhomogeneities. The scale bar is 10 nm. **c)** Elemental analysis TEM (EDX) of a $PbI_2$ flake. Lead is shown in purple. Iodine is shown in blue. The scale bar is 50 nm. Regions of lead buildup can be observed. **d)** PL versus power for the 4-layer heterostructure region, excited with 488 nm. Power increases from 15 µW to 160 µW. The endpoint colors corresponding to 15 and 160 µW are indicated in the legend, and intermediate traces correspond to successive 10% increments of the applied laser power. Background is subtracted. The inset shows the differential heterostructure PL, defined as the PL from the $PbI_2$ region subtracted from the 4-layer heterostructure region. Subtracting the STE contribution recovers the defect PL. **e)** Power plot of the intensities of the $V_B^-$ peak in heterostructure vs in neutron-irradiated hBN. The standard deviations of measurements are shown as bars while the standard deviation of the fit is shown in shading. **f)** Difference between $V_B^-$ ensemble emission coupled and uncoupled to a $PbI_2$ donor layer, plotted against the baseline uncoupled defect luminescence. The standard deviations of measurements are shown as bars while the standard deviation of the fit is shown in shading. Both e and f use a laser with 488 nm excitation energy. **g)** PL intensity of the low-energy defect spectral region versus temperature, excited with 460 nm.

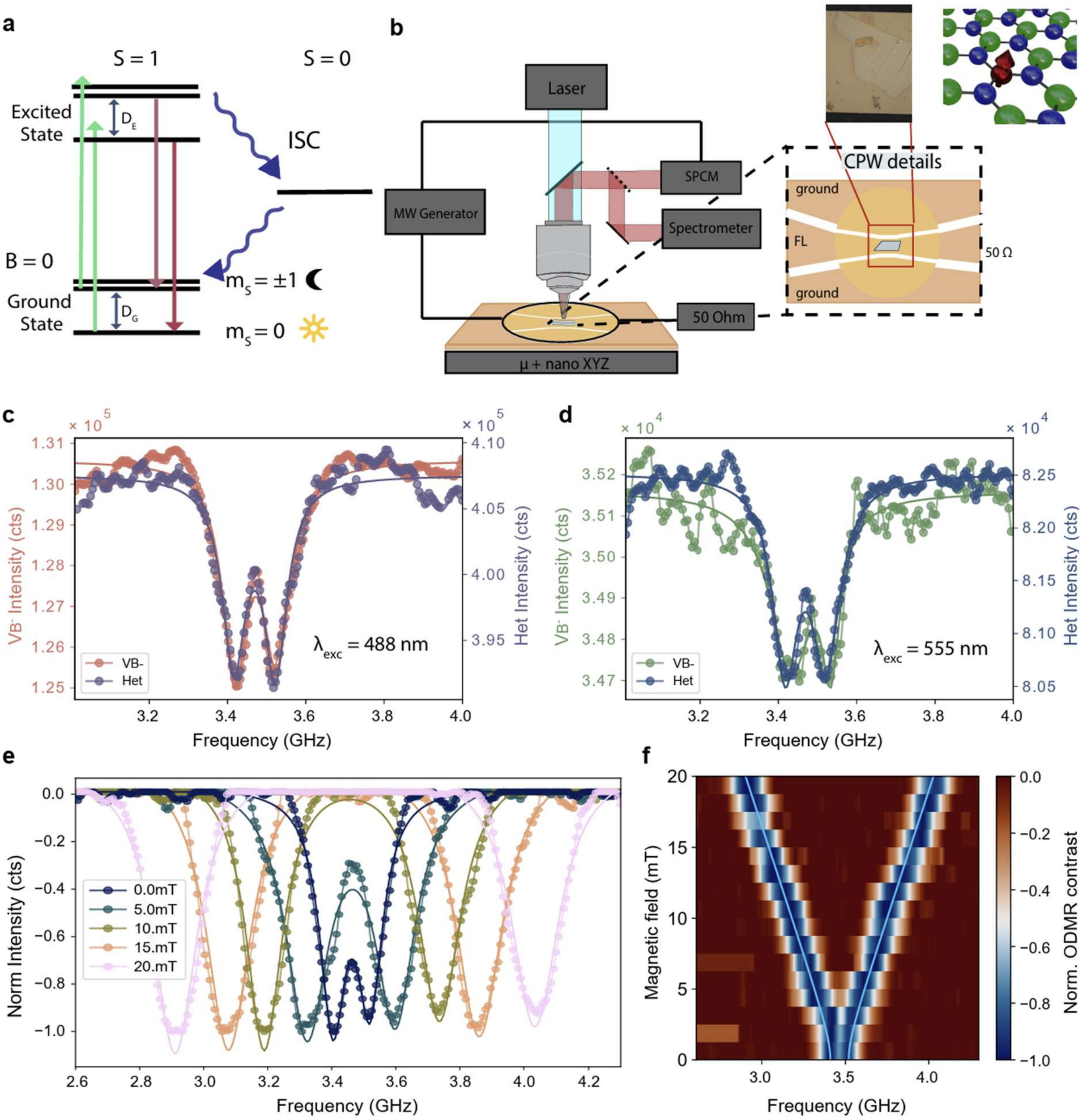

**Figure 5 ODMR and Sensing Magnetic Fields: a)** Energy level diagram for $V_B^-$ and schematic illustration of the readout by ODMR. The system was excited off-resonantly via a laser (green arrow). ODMR contrast is governed by differing rates of radiative decay (red arrows) vs nonradiative decay via intersystem crossing (ISC, purple arrow) to the meta-stable (MS) singlet state from the excited state to ground state. Dips in photoluminescence are observed at the frequency corresponding to the energy difference between $m_s = 0$ and $m_s = \pm 1$ ($D_{gs}$). **b)** Schematic illustration of the ODMR setup. A coplanar waveguide (CPW) is placed in the center of the path with the heterostructure placed onto the feedline (antenna). This CPW is wire-bonded to a PCB, which interfaces with the signal generator (SG) and 50 Ohm impedance termination. The sample is excited using a laser, and emission is detected by the APD, which is triggered by the change in frequency from the SG**. c)** ODMR spectra when the $V_B^-$ (left, pink) vs the heterostructure (right, purple) are excited by a 488 nm laser. **d)** ODMR spectra when the $V_B^-$ (left, green) vs the heterostructure (right, blue) are excited by a 555 nm laser. **e)** Normalized ODMR of the heterostructure with varying magnetic fields. **f)** Data points from e, as well as additional points, are plotted as the magnetic field versus resonance frequency.

# SUPPLEMENTARY INFORMATION FOR

## A van der Waals photon-ratchet for enhancing spin-defect emission and quantum sensing

E. Mayner[1*], Y. Zhumagulov[2], C. de Giorgio[3], F. Chu[1], P. P. Swain[4], G.E. Fantner[4], A. Kis[3], O. V. Yazyev[2], A. Radenovic[1*]

[1]LBEN, Bioengineering Institute, École Polytechnique Fédérale de Lausanne (EPFL), CH-1015 Lausanne, Switzerland.

[2]C3MP, Institute of Physics, École Polytechnique Fédérale de Lausanne (EPFL), CH-1015 Lausanne, Switzerland

[3]LANES, Electrical Engineering Institute, École Polytechnique Fédérale de Lausanne (EPFL), CH-1015 Lausanne, Switzerland

[4]LBNI, Bioengineering Institute, École Polytechnique Fédérale de Lausanne (EPFL), CH-1015 Lausanne, Switzerland.

[*]Email: eveline.mayner@epfl.ch, aleksandra.radenovic@epfl.ch

# 1 CONTENTS

# 2 AB INITIO DFT

## 2.1 ELECTRONIC-STRUCTURE CALCULATIONS

As a starting point for our theoretical analysis, we performed DFT calculations of a free-standing hBN monolayer and an hBN-$PbI_2$ heterostructure supercell containing the negatively charged boron vacancy, $V_B^-$ (SI Fig. 1). The heterostructure is modeled by a $(\sqrt{28}\times\sqrt{28})R19.1°$ hBN supercell matched to a (3×3) $PbI_2$ supercell: a common hexagonal cell with a = 13.25 Å containing 28 hBN formula units (one boron vacancy gives the $V_B^-$ center) and 9 $PbI_2$ formula units. The hBN layer hosting the defect is kept at the experimental lattice constant, a = 2.504 Å, and the residual mismatch is accommodated by the $PbI_2$ layer, which is compressed by 0.75% to 4.417 Å. Calculations were carried out using the GPAW code[1] in plane-wave mode, with a kinetic-energy cutoff of 500 eV and the local-density approximation (LDA) for exchange–correlation potential. The defect-containing cell carries a total charge of −1, in correspondence of the charge state of $V_B^-$. Since the plane-wave basis is periodic in all three directions, the layers were separated from their periodic images along the *z* direction by 10 Å of vacuum. Structural relaxation was performed with the LBFGS algorithm, relaxing the in-plane lattice parameters and all ionic positions while keeping the out-of-plane lattice parameter and cell angles fixed, until the maximum residual force was below $10^{-3}$ eV Å$^{-1}$.

The resulting Kohn–Sham eigenvalues $\varepsilon_n(\mathbf{k})$ and cell-periodic Bloch states $u_{n\boldsymbol{k}}$ provide the microscopic input for the optical response, including band-energy differences, occupations, and optical matrix elements that enter the perturbative (independent-particle) description of light–matter coupling used below.

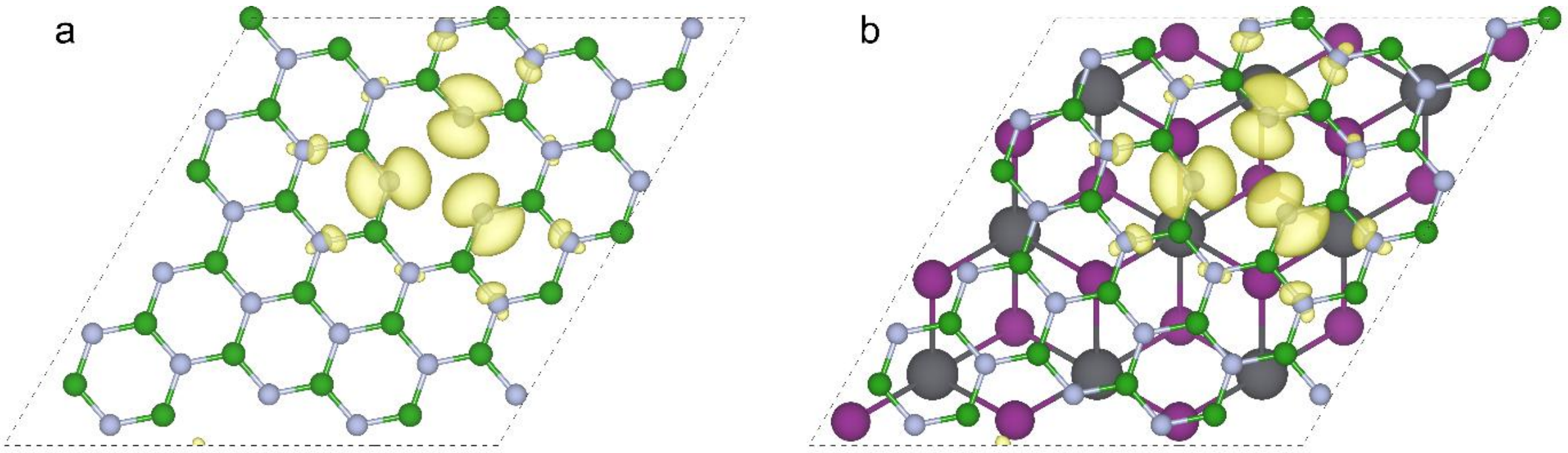

**SI Figure 1 Calculated Spin Densities**: Spin density distribution at the $V_B^-$ defect in **a)** the isolated hBN and **b)** the hBN-$PbI_2$ heterostructure.

## 2.2 SECOND-ORDER SUSCEPTIBILITY AND PUMP-INDUCED EMISSION

To quantify how the $PbI_2$ substrate modifies the $V_B^-$ optical response, we compute the second-order susceptibility tensor $\chi(2)$ within the independent-particle approximation (IPA) using the length-gauge Aversa–Sipe formalism.[2,3] We connect $\chi(2)$ to the emission generated under monochromatic laser pumping in the spectral vicinity of the $V_B^-$ defect line. In this framework, the calculated substrate-induced changes in emission can be traced back to modifications of the underlying electronic structure and optical matrix elements.

We consider a driving field with complex frequency components $E_j(\omega_1)$ and the induced second-order polarization. With two frequency arguments $+\omega_1$ and $-\omega_2$ in the susceptibility, the emitted/observed frequency is $\Omega = \omega_1 - \omega_2$. The second-order polarization at frequency $-\Omega$ reads

$$P_i^{(2)}(-\Omega) = \sum_{jk} \chi_{ijk}^{(2)}(-\Omega; \omega_1, -\omega_2) E_j(\omega_1) E_k^*(\omega_2). \tag{1}$$

In what follows, the emission signal is taken to scale with the magnitude of the nonlinear polarization at the emitted frequency, $I(\Omega) \propto \left|\mathbf{P}^{(2)}(\Omega)\right|^2$ (up to geometry- and collection-dependent prefactors).

We work with Bloch states $u_{nk}$ of band $n$ at wavevector $\mathbf{k}$, with energies $\varepsilon_n(\boldsymbol{k})$ and occupations $f_n$. Define

$$\omega_{mn}(\mathbf{k}) \equiv \frac{\varepsilon_m(\mathbf{k}) - \varepsilon_n(\mathbf{k})}{\hbar}, f_{nm} \equiv f_n - f_m. \tag{2}$$

Interband position matrix elements and the generalized (covariant) $k$-derivative are

$$\begin{aligned} r_{nm}^a(\boldsymbol{k}) &\equiv i\langle u_{n\boldsymbol{k}} | \partial_{k_a} | u_{m\boldsymbol{k}} \rangle (n \neq m), \\ (r_{nm}^a)_{;b} &\equiv \partial_{k_b} r_{nm}^a - i(\xi_{nn}^b - \xi_{mm}^b) r_{nm}^a \end{aligned} \tag{3}$$

with Berry connections $\xi_{nn}^b = i\langle u_{n\boldsymbol{k}} | \partial_{k_b} u_{n\boldsymbol{k}} \rangle$.

Within IPA and the length gauge, the quadratic susceptibility at mixed frequencies $(\Omega; \omega_1, \omega_2)$ can be written as

$$\chi^{(2),abc}(-\Omega;\omega_1,-\omega_2)=\frac{e^3}{\hbar^2}\int\frac{d^3k}{(2\pi)^3}\left[\chi_{2B}^{abc}(\boldsymbol{k})+\chi_{3B}^{abc}(\boldsymbol{k})\right],\tag{4}$$

where $a,b,c\in\{x,y,z\}$, and the result is decomposed into a two-band and a three-band contribution:

$$\begin{gathered}\chi_{2B}^{abc}(\mathbf{k})=\sum_{n\neq m}\frac{f_{nm}r_{nm}^{a}}{\omega_{mn}-\Omega+i\eta}\\ \left(\frac{\left(r_{mn}^{b}\right)_{;c}}{\omega_{mn}-\omega_1+i\eta}+\frac{(r_{mn}^{c})_{;b}}{\omega_{mn}+\omega_2+i\eta}\right),\\ \chi_{3B}^{abc}(\boldsymbol{k})=\sum_{\substack{n\neq m\\ \ell}}\frac{r_{nm}^{a}r_{m\ell}^{b}r_{\ell n}^{c}}{\omega_{\ell n}-\omega_2+i\eta}\\ \left(\frac{2f_{nm}}{\omega_{mn}-\Omega+i\eta}-\frac{f_{n\ell}}{\omega_{\ell n}-\omega_1+i\eta}-\frac{f_{m\ell}}{\omega_{m\ell}+\omega_2+i\eta}\right).\end{gathered}\tag{5}$$

These equations reduce to the standard SHG expression for $\Omega=2\omega$ with $\omega_1=\omega_2=\omega$, and to optical rectification for $\Omega=0$ with $\omega_1=\omega_2=\omega$.

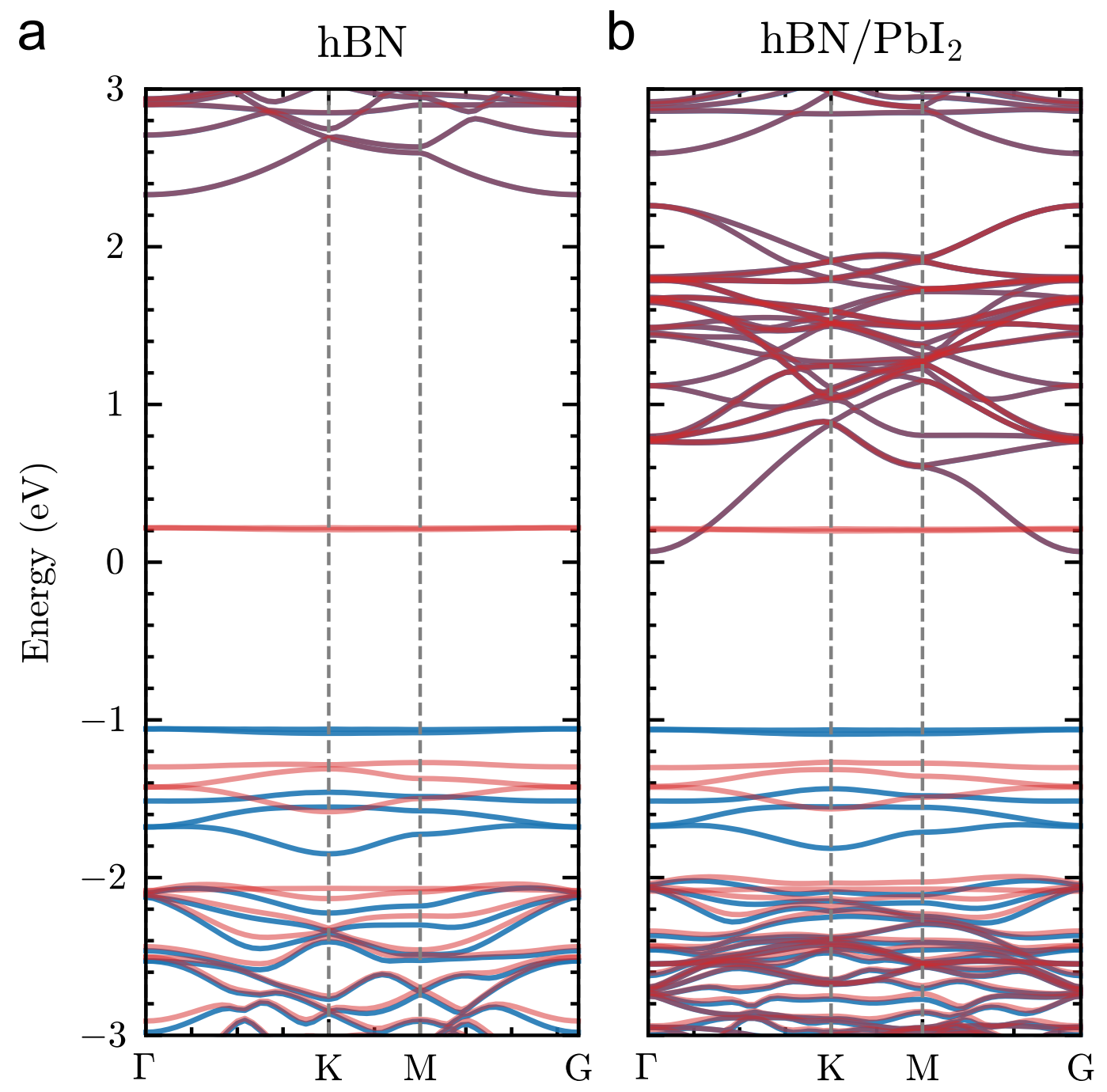

**SI Figure 2 Spin Band Diagram**: Spin-resolved band structures of $V_B^-$ in hBN calculated **a)** in the absence and **b)** in the presence of $PbI_2$. Spin up and spin down are shown in blue and red, respectively. 0 is the Fermi level at 300 K.

# 3 Extended Materials Characterization

The following section provides additional details on the characterization of the materials used in this study.

## 3.1 Additional Optical Characterization

The photoluminescence excitation spectra (PLE) of the four-layer region of the heterostructure from Main Text Figure 5 was recorded with excitation wavelengths spanning from 400 to 740 nm (i.e., below the PbI2 bandgap to near the defect peak). Data is shown in SI Fig. 3. The confocal setup described in Methods was used in combination with a SuperK Fianium supercontinuum laser and an additional alternate path (flip mirror), a power meter, to ensure constant power throughout the sweep. The PLE shows spikes in emission from the defect, concurring with theoretical absorption and PL spectra (Main Text Figure 2c, d). Significant absorption is also observed when exciting with energies near but lower than the $PbI_2$ bandgap, indicating the potential bandgap renormalization discussed in the main text. A 775 nm long-pass (LP) filter was used to filter out the majority of the tail of the $PbI_2$ STE in order to detect emission mainly from the $V_B^-$.

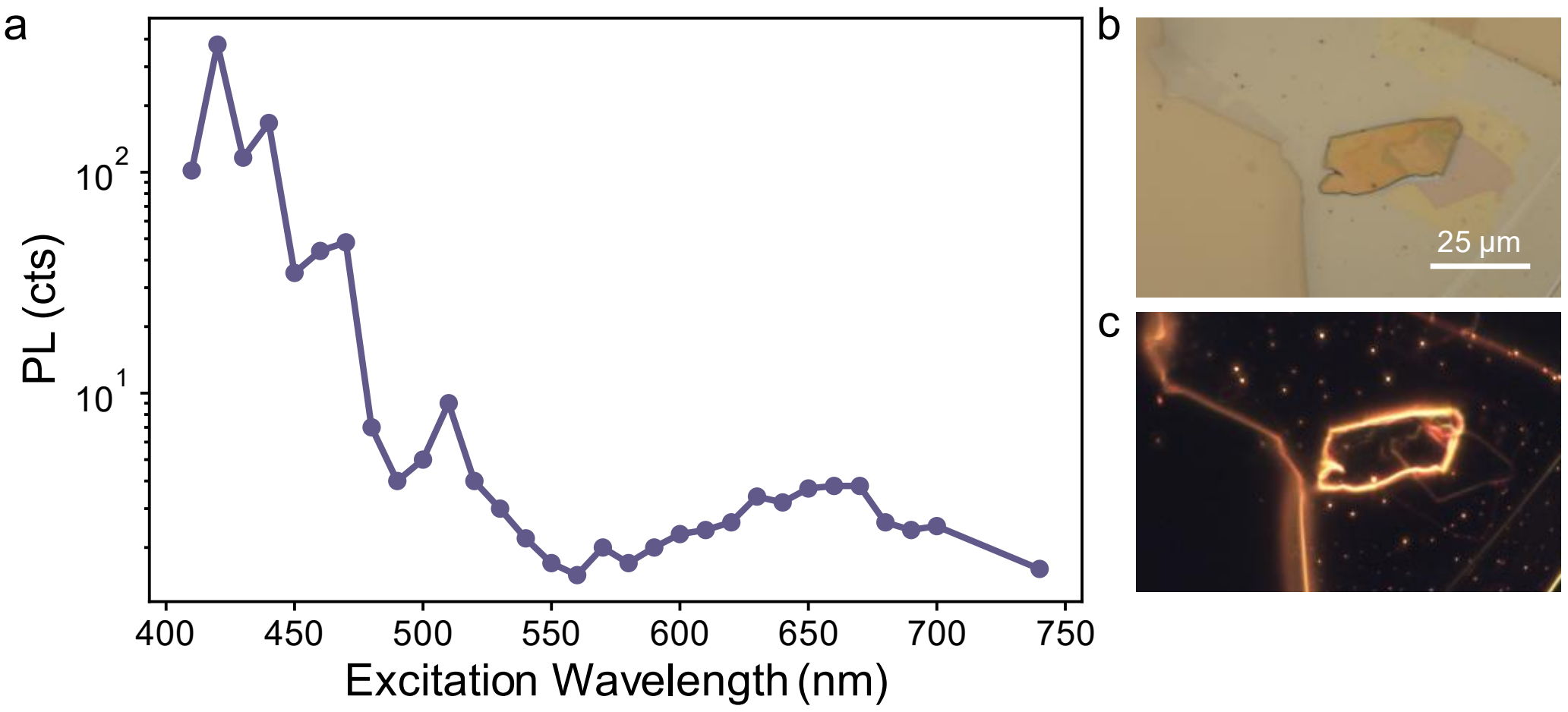

**SI Figure 3 PLE of Heterostructure**: **a)** PLE spectra from the four-layer region of the heterostructure shown in white light **b)** and darkfield **c)**. This is the same heterostructure as in Main Text Figure 5.

The differential PL from the heterostructure and $V_B^-$ regions of the heterostructure used in RT measurements in Main Text Figure 3 shows an enhancement factor of 15× (SI Fig. 4), significantly lower than the 45× observed at 4K. Data from another heterostructure analyzed at room temperature (RT) is shown in SI Fig. 5. Instead of differential PL, the fit of the three peaks (FE, STE, $V_B^-$) within the heterostructure spectra is shown.

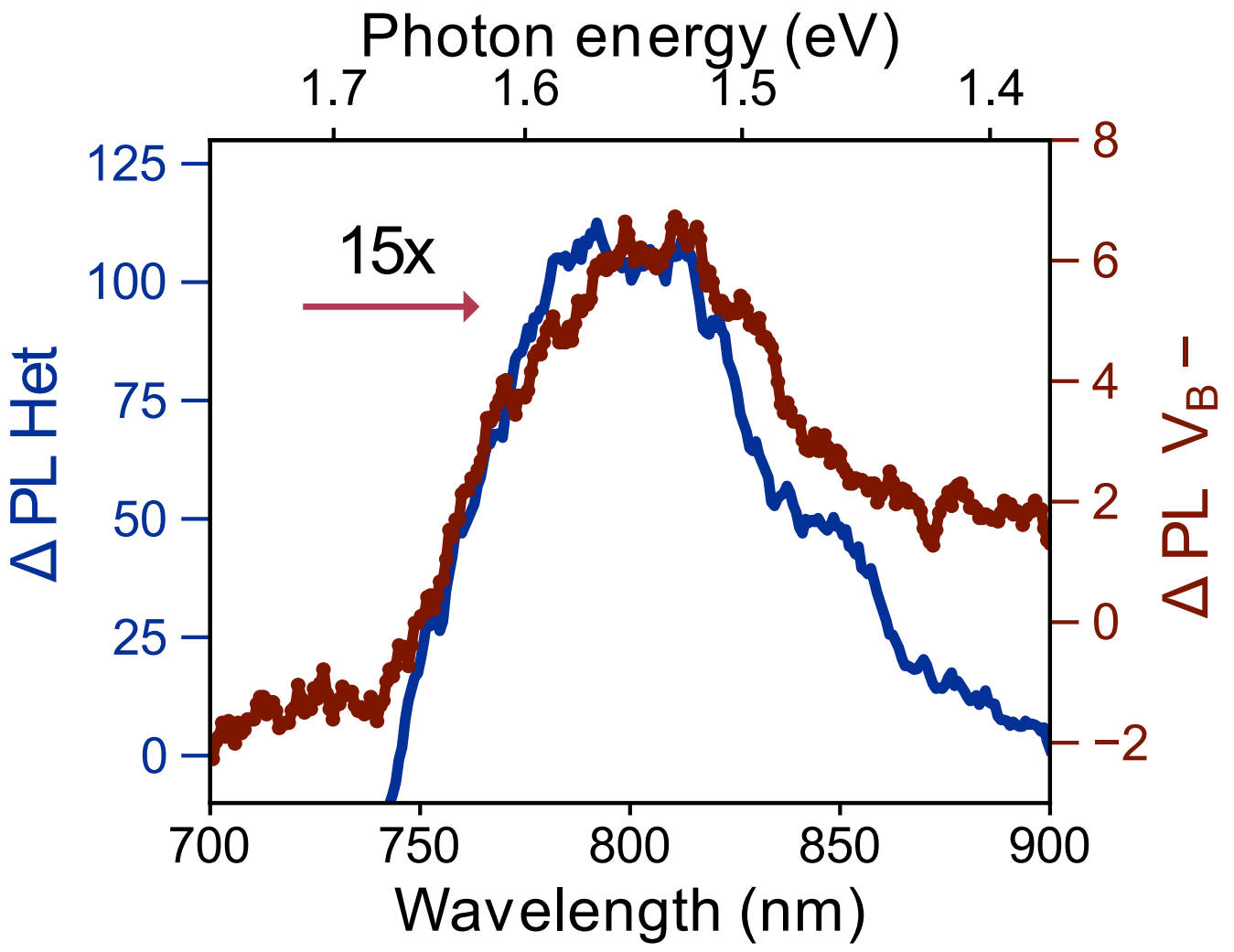

**SI Figure 4 Room Temperature Differential PL**: The differential PL spectra for emitter and PL in Main Text Figure 3. $PbI_2$ spectrum was subtracted from the heterostructure spectrum. The background spectrum was subtracted from the $V_B^-$ spectrum. Data is plotted on different y-scales.

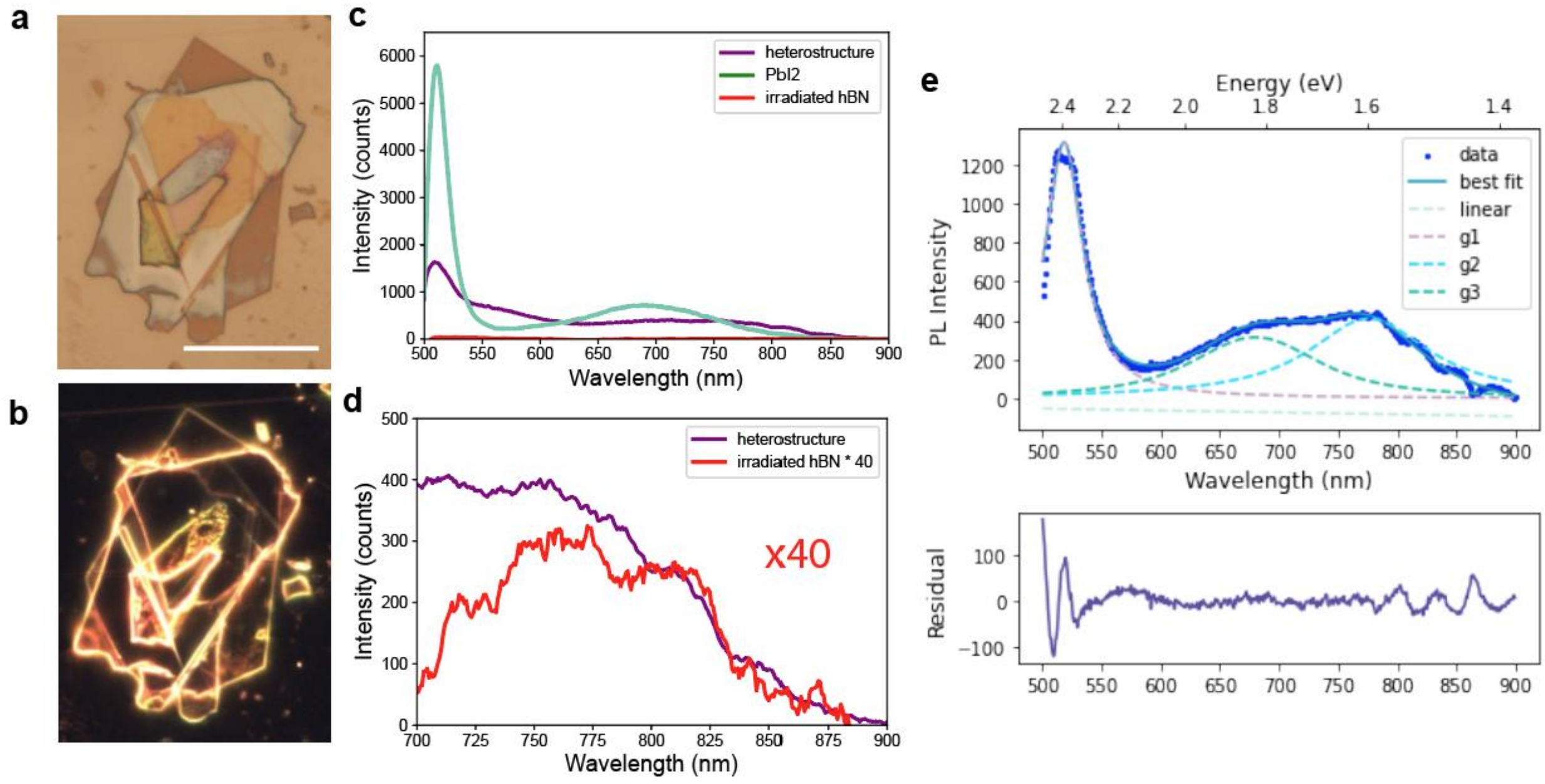

**SI Figure 5 Additional Example of hBN-$PbI_2$ Heterostructure**: **a)** White-light image of another example of the fabricated hBN-$PbI_2$ heterostructure. The scale bar is 50 µm. **b)** Dark-field image of the same heterostructure. **c)** PL spectra taken from the three regions of interest (ROIs) of the heterostructure, as in the Main Text: heterostructure region (all four layers), $PbI_2$ ($PbI_2$ plus hBN capping layers), and the irradiated hBN (irradiated hBN plus capping layers). The peak that appears between 550-600 nm is attributed to visible range defects in hBN, often attributed to carbon.[4,5] They are often brighter than the $V_B^-$ but could also be enhanced by the $PbI_2$ in this case. **d)** Detailed spectrum from the heterostructure and the irradiated hBN, with the irradiated hBN spectrum multiplied by 40 for scaled matching. **e)** Three peak fitting of the heterostructure with a linear background. The bottom graph is the residual function of the best fit.

At 4K, the lifetime of the coupled structure is longer than that of the neutron irradiated hBN alone (SI Fig. 6a). At room temperature, the emission from the coupled region exhibits a shorter lifetime as compared to the isolated neutron irradiated hBN (SI Fig. 6c). A 775 LP filter was used in both cases. However, at room temperature the broadband red emission from $PbI_2$ overlaps more strongly with the $V_B^-$ defect-emission spectral window, which may cause the lifetime decrease. At low temperatures the intrinsic emission appears to increase in the coupled structure. This may be due to a delayed feeding of $V_B^-$ from the $PbI_2$ exciton reservoir, although residual $PbI_2$-associated emission cannot be excluded. The decays were fitted with instrument response function (IRF)-deconvoluted biexponential decay functions,

$$I(t) = \sum_i^2 A_i e^{-t/\tau_i} + C \,.$$

A bi-exponential model has been used previously to describe $V_B^-$ PL decay in hBN alone, where the measured lifetime reflects various recombination channels.[6,7] In the heterostructure, such behaviour may arise from overlapping emissive contributions from $V_B^-$ and residual $PbI_2$ self-trapped-exciton emission. The bi-exponential model captures the dominant decay behavior but does not uniquely assign all long-time residual signal to a microscopic channel. Therefore, individual fit components are not interpreted, and the intensity-weighted average lifetime is used as a comparison metric. The lifetime values reported in the legends and discussed in the text correspond to the intensity-weighted average lifetime,

$$\tau_{av} = \frac{\sum_i^2 A_i \tau_i^2}{\sum_i^2 A_i \tau_i},$$

rather than to an individual fit component. However, the individual lifetime components and their amplitudes are reported in **Table 1 Lifetime Components from SI Figure 6**.

Exciting with 460 nm and *without* a 775 nm long-pass filter in the emission path, the fluorescence is expected to be dominated by the free exciton emission of $PbI_2$. This data is fit with a deconvoluted bi-exponential decay to account for the FE as multiple emissive/recombination environments. In the coupled structure, a slight reduction of the lifetime is observed compared to that of $PbI_2$ alone. The lifetime shortening of the energy donor is characteristic of energy-transfer processes. However, the effect is too weak to draw firm conclusions.

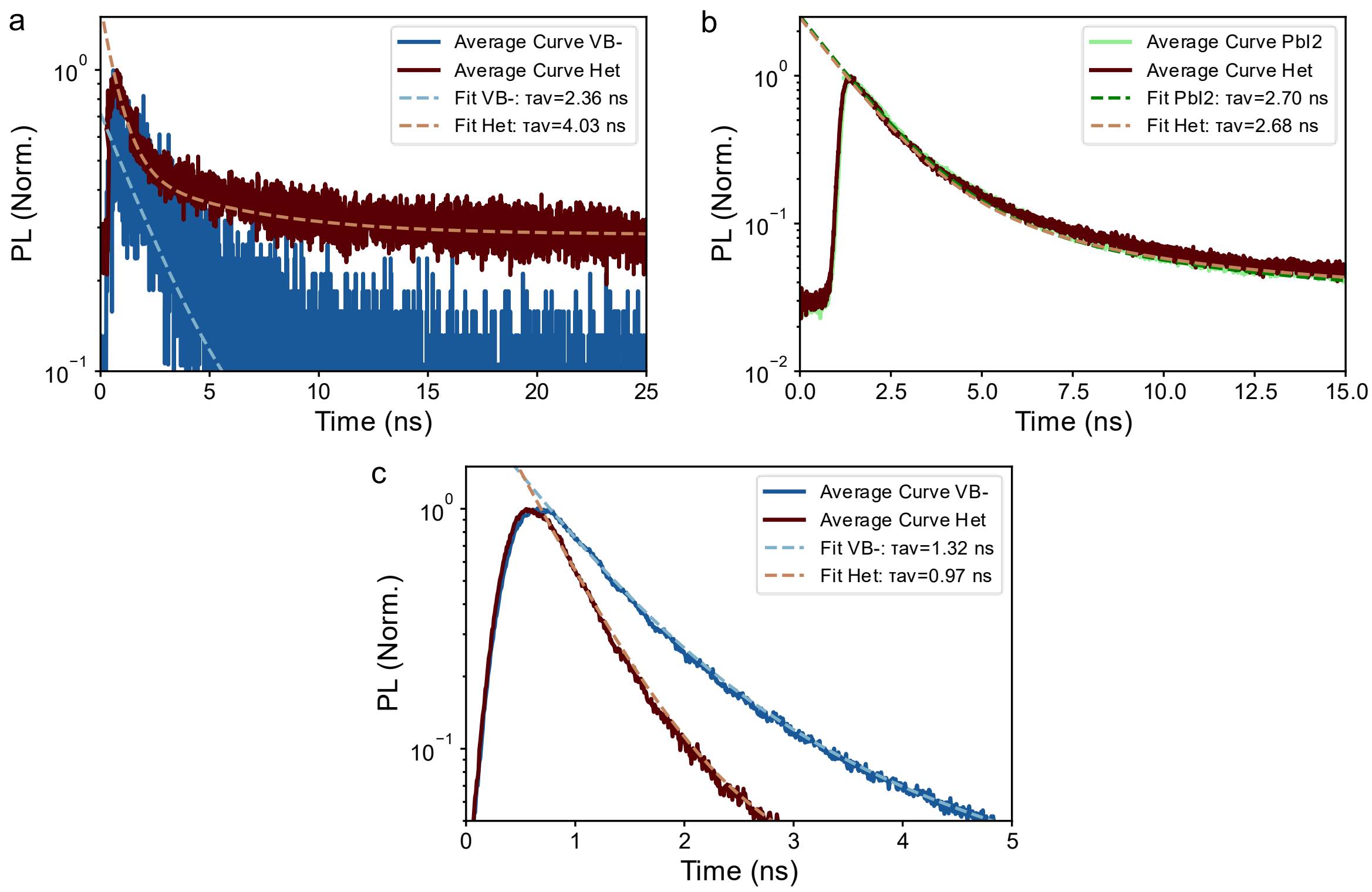

**SI Figure 6 Lifetimes with 460/555 nm Excitation at 4K and Room Temperature**: **a)** Comparison of the time-resolved PL decay of the heterostructure and neutron-irradiated hBN under 460 nm excitation with a 775 nm long-pass filter. Data recorded at 4K. The PL intensity is normalized, and traces were recorded with a 250 GHz sampling rate. The heterostructure (coupled structure, red) and the $V_B^-$ region (uncoupled, blue) are both deconvoluted with the IRF and fit with a bi-exponential decay. The fits are shown and the legend reports the corresponding intensity-weighted average lifetimes. All lifetimes and their weights are reported in Table 1. **b)** PL decay of the heterostructure and $PbI_2$ under 460 nm excitation without the 775 nm long-pass filter. The PL intensity is normalized and deconvoluted bi-exponential fits along with the derived intensity-weighted average lifetimes are displayed. Data recorded at 4K. A 250 GHz sampling rate was used. **c)** Lifetimes of the defect (observed using a 775 LP filter) in the coupled (red) vs uncoupled (blue) structures, excited with 555 nm, at room temperature Both are fit with IRF-deconvoluted bi-exponential decay functions and the intensity weighted average is reported in the legend.

**Table 1 Lifetime Components from SI Figure 6**

| **Sample** | $\tau_1(A_1)$ | $\tau_2(A_2)$ | $\tau_{av}$ |
|---|---|---|---|
| **4K $V_B^-$ 775 LP (6a)** | 2.0 (0.08) | 2.41 (0.45) | 2.36 |
| **4K heterostructure 775 LP (6a)** | 0.77 (0.51) | 5.39 (0.17) | 4.03 |
| **4K $PbI_2$ (6b)** | 1.03 (0.71) | 2.76 (0.32) | 2.70 |
| **4K heterostructure (6b)** | 0.99 (0.68) | 2.66 (0.33) | 2.68 |
| **RT $V_B^-$ 775 LP (6c)** | 0.688 (0.90) | 2.42 (0.21) | 1.32 |
| **RT heterostructure 775 LP (6c)** | 0.48 (1.15) | 1.59 (0.098) | 0.97 |

## 3.2 $PbI_2$ STE

The energy landscape of the self-trapped exciton (STE) formation is illustrated in SI Fig. 7a. The STE involves optical excitation of $PbI_2$ and the formation of a free exciton that descends into a local energy minimum caused by a lattice distortion within the material. Once trapped, this exciton exhibits radiative decay at a lower energy, leading to the characteristic redshifted emission relative to the FE. The broad spectral profile of the STE is attributed to multiple transitions across vibronic states. $PbI_2$ flakes were also imaged when capped with hBN and when uncapped to determine the effect of capping on structure integrity.

The PL of both uncapped and capped $PbI_2$ was recorded with increasing laser power (here, in percent of 3.2 kW $cm^{-2}$). In the uncapped configuration, FE decreases in intensity despite increasing power, while the STE increases in intensity and eventually plateaus (SI Fig. 7b). The decrease in FE peak and increase in STE peak indicate the material is degrading with higher excitation powers. This degradation seems to be inhibited when $PbI_2$ is capped (SI Fig. 7c), suggesting that the degradation mechanism involves oxygen. In SI Fig. 7 panels d-e, one can see the rapid changes to the material under the electron beam used in Transmission Electron Microscopy (TEM) with 200 kV high tension. Attention should especially be paid to the bottom right corner of the image, where layers even appear to peel. Both measurements suggest structural instability upon imaging, either with an electron beam or with a laser.

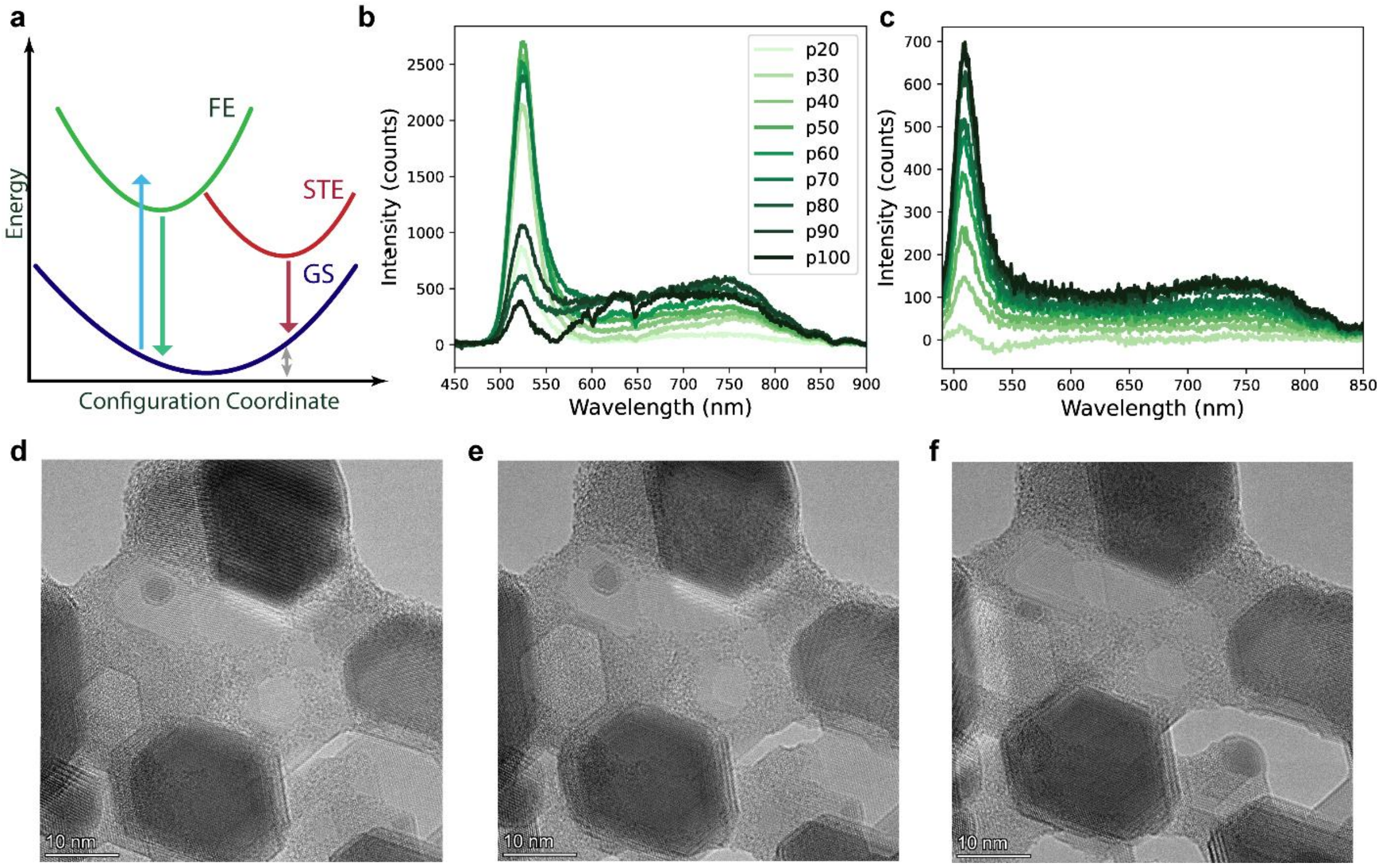

**SI Figure 7 Lead Iodide and STE**: **a)** Drawing of the STE formation in $PbI_2$ in the Franck-Condon picture. The FE line represents the free exciton state, the STE line represents the self-trapped exciton state, and the GS line represents the ground state. The blue arrow represents the excitation. The green arrow represents the FE emission. The red arrow represents the STE emission, and the gray double-sided arrow represents the lattice distortion energy. Lead iodide photoluminescence with background (off-flake, on-coverslip) subtracted **b)** uncapped and **c)** capped. Panels **d-f** show high-resolution TEM images taken continuously with 1 second dwell times, indicating beam damage. Beam voltage was 200kV.

## 3.3 Power Dependent Spectral Analysis

SI Fig. 8a and 8b show the differential PL of the STE and $V_B^-$ regions for the heterostructure in Main Text Figure 4. The STE and $V_B^-$ peaks both increase as a function of the 488 nm excitation power. The $V_B^-$ peak intensity increases approximately 14× faster in the heterostructure than in neutron-irradiated hBN alone. The intensity analysis as a function of power in the Main Text and here in the SI are done using the integrated intensity. Integrated intensities are calculated by the areas under the PL fit (examples shown in SI Fig 8 d-f). We also observed a power law near linear scaling for the $V_B^-$ peak in Main Text Figure 4, even when coupled with $PbI_2$. Given the effectively unlimited $PbI_2$ exciton reservoir, this behaviour supports a transfer-limited regime, where ET can be considered inefficient and in which a substantial fraction of donor excitons recombine through competing pathways. Consider the intensity ($I_{V_B-}$) as a function of the rate of the ET ($\eta_{ET}$) and the number of free excitons in $PbI_2$ ($G_{PbI_2}$):

$$I_{V_B-} \propto \eta_{ET} * G_{PbI_2}. \tag{6}$$

Since the intensity of the emission from the defects was still limited by laser power ($P_{exc}$)

$$I_{V_B-} \propto P_{exc}, \tag{7}$$

we can consider that the product of the efficiency and the number of excitons is still less than the number of defect sites available ($N_{V_B-}$):

$$\eta_{ET} * G_{PbI_2} < N_{V_B-} \tag{8}$$

The experiment was done with a relatively thick hBN flake (>10 nm) and therefore would be more informative of the rate of transfer and the maximal PL enhancement factor under conditions where $N_{V_B-}$ is a limiting factor, such as in a thin or even monolayer flake.

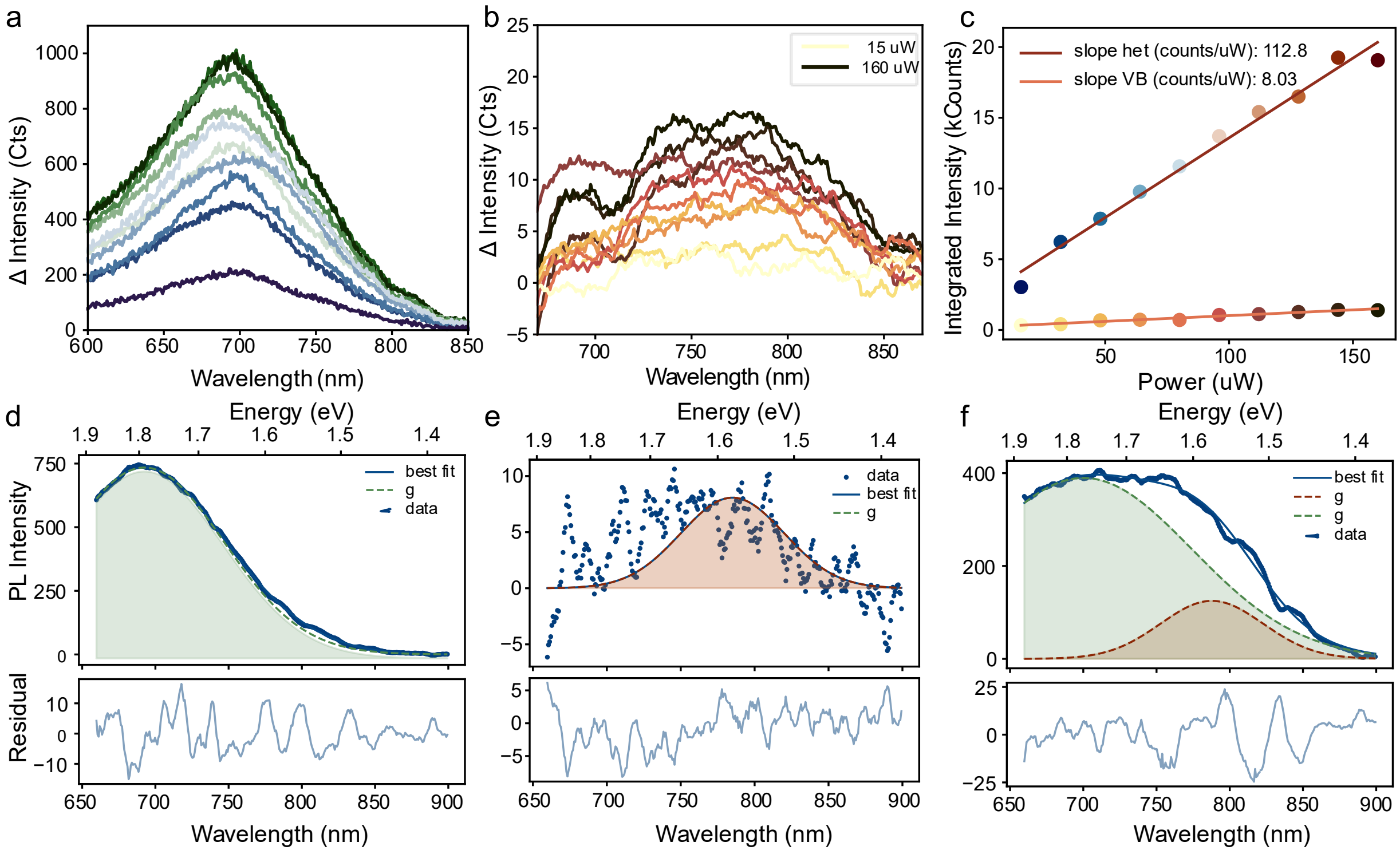

**SI Figure 8 Details on Power Analysis**: **a)** Spectra from Figure 4a rescaled to show the emission from the STE peak as a function of the 488 nm excitation laser power. **b)** The differential intensity (background spectrum subtracted) of the $V_B^-$ region of the heterostructure as a function of the 488 nm excitation laser power. **c)** Linear plot of integrated intensity vs laser power for the heterostructure and $V_B^-$ regions. Best fits and corresponding slopes are in the legend. **d)** One Gaussian peak fitting with only $PbI_2$ present, fitting the STE. **e)** Example Gaussian fitting of the neutron irradiated section of the heterostructure, with background subtracted. Peak fitting is limited to a center of 785 nm or above. **f)** Two Gaussian peak fitting of the heterostructure region where all four layers are present. The second peak is also constrained to a center of 785 nm or above.

## 3.4 AFM AND LAYER THICKNESSES

To complement the optical measurements, we next report the layer thicknesses of representative devices. Atomic force microscopy (AFM) height profiles and the corresponding layer thicknesses are shown in SI Fig. 9 and summarized in SI Table 2 AFM Characterization. The measured devices contain irradiated-hBN layers with thicknesses of several tens of nanometers and $PbI_2$ layers of approximately one hundred nanometers or greater. Because the devices differ in defect density, initial brightness, $PbI_2$ morphology and broadband-emission background, these measurements are provided as structural characterization rather than as a controlled thickness series. Accordingly, no monotonic relationship between layer thickness and PL or ODMR enhancement is inferred from this dataset.

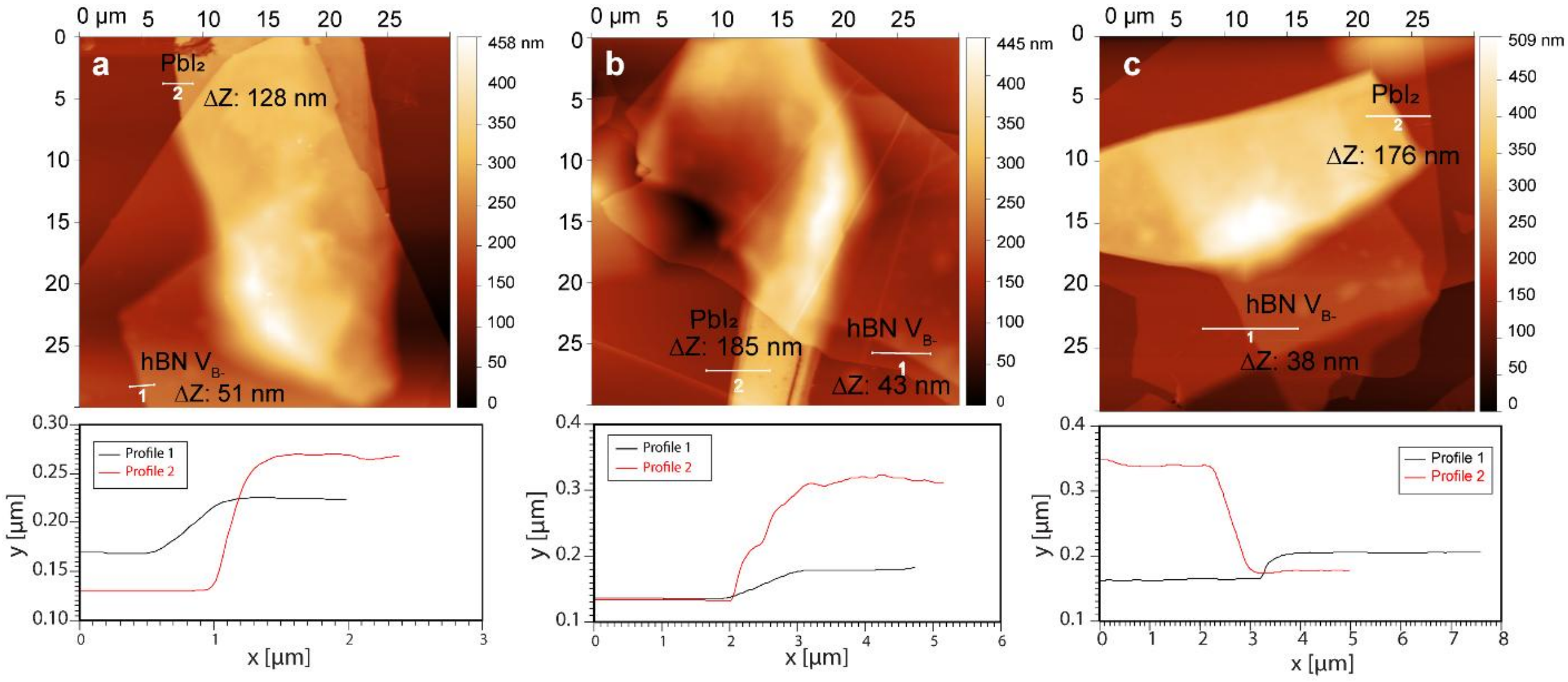

**SI Figure 9 Atomic Force Microscopy (AFM)**: **a)** Sample 2 (Main Text Figure 5). **b)** Sample 3 **c)** Sample 4.

**Table 2 AFM Characterization from SI Figure 9**

| **Device** | **Corresponding figure/data** | **AFM Scan** | **Thickness $V_B^-$ (nm)** | **Thickness $PbI_2$ (nm)** |
|---|---|---|---|---|
| **S1** | Main Fig. 1; SI Fig. 14 | N/A | - | - |
| **S2** | Main Fig. 5; SI Fig. 12-13 | SI Figure 9a | 51 | 128 |
| **S3** | -- | SI Figure 9b | 43 | 185 |
| **S4** | SI Fig. 15 | SI Figure 9c | 38 | 176 |

## 3.5 SPATIAL VARIATION OF $PbI_2$ BROADBAND EMISSION

Because $PbI_2$ emission is sensitive to local structural and chemical environment, we examined individual spectra from within the heterostructure from Main Text Fig. 3. SI Fig. 10c shows individual spectra from different locations within the $PbI_2$ flake, which are marked with colored stars on the heat map in SI Fig. 10b (reproduced from Main Text Fig. 3 for clarity). Strain or twisting within the layers of the $PbI_2$ flake can cause an additional, red-shifted peak, to appear. In this heterostructure, strained regions within the $PbI_2$ flake exhibit an additional red-shifted broadband emission component, which extends into the 780–820 nm detection window. This behavior is consistent with local variations in exciton–phonon coupling, stacking/twist, defect density, or Pb–I coordination. Similar stress-induced transitions between distinct STE emission states have recently been reported in another semiconductor system, where nonuniform/deviatoric stress was shown to induce an additional STE state and modify the carrier relaxation pathway.[8] In $PbI_2$ nanoclusters, time-resolved photoluminescence has shown the formation of self-trapped excitons, while twisted $PbI_2$/$PbI_2$ heterostructures exhibit twist-angle-triggered exciton–phonon coupling that introduces self-trapped levels within the $PbI_2$ bandgap and extends the optical response toward the near-infrared.[9,10] At the same time, broad red-shifted emission in lead-iodide-based layered materials has also been attributed to extrinsic in-gap or defect-related states, including iodine-vacancy-related states, rather than purely intrinsic STE recombination.[11]

These measurements establish that the off-overlap signal in Main Text Fig. 3 likely originates from a spatially varying $PbI_2$-associated broadband component. Its implications for ODMR sensitivity are discussed in section 4.2.

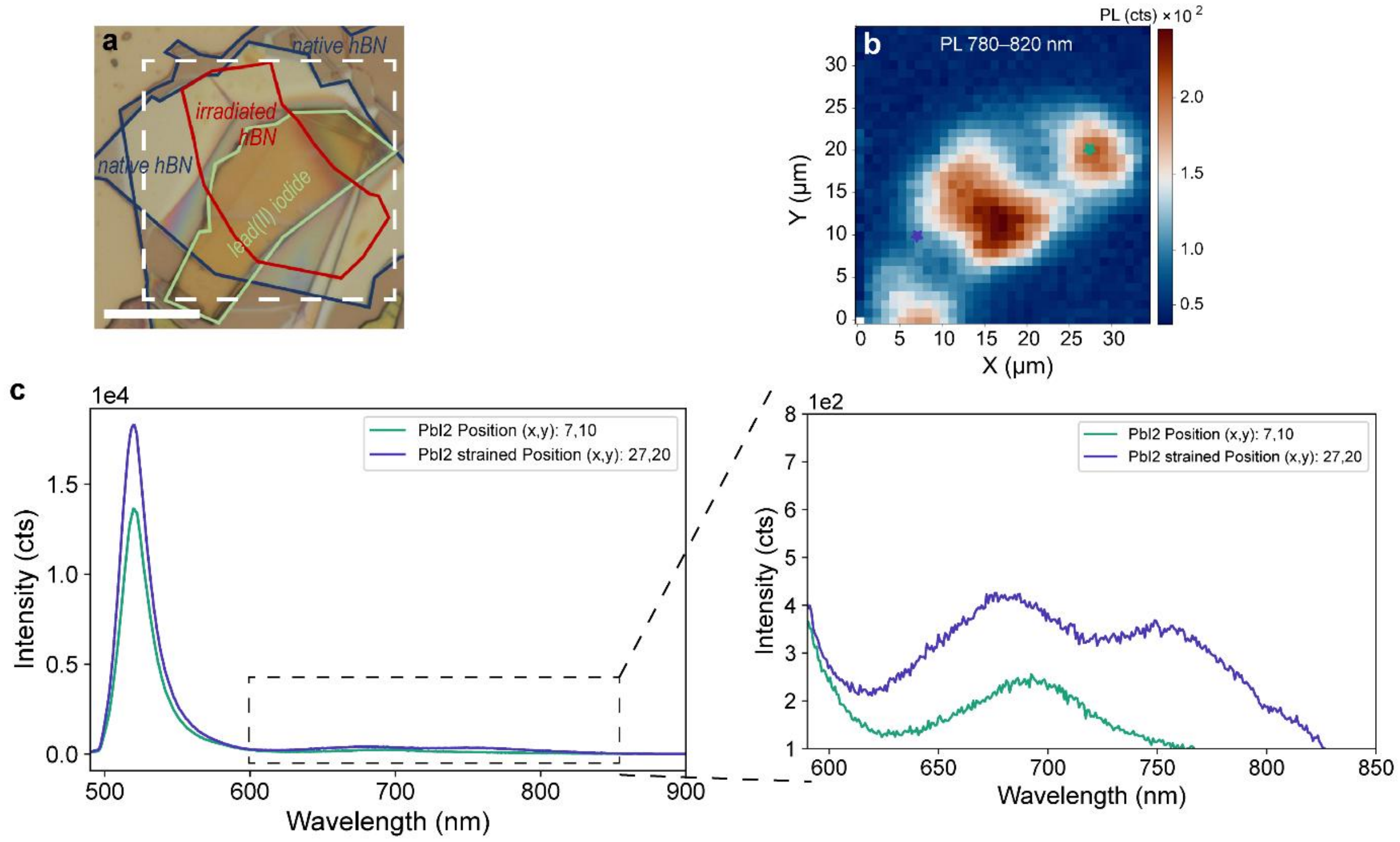

**SI Figure 10 Local Spectra from Heterostructure with Strained $PbI_2$: a)** Optical image of the flake shown in Main Text Fig. 3, included for reference. **b)** PL intensity map integrated from 780–820 nm, with colored stars marking the positions in the $PbI_2$ flake where spectra were extracted. **c)** Individual spectra from the marked positions in the $PbI_2$ flake as well as from other regions of the heterostructure. The color of the star corresponds to the color of the plotted spectra. The zoomed spectral region highlights the overlap between $PbI_2$-associated broadband emission and the $V_B^-$ emission window.

# 4 Optically Detected Magnetic Resonance Sensitivity Details

This section reports the additional details on sensitivity calculations and further experiments on the sensitivity vs power and sensitivity over time.

## 4.1 Detailed ODMR Characterization

**Table 3 Fit parameters from Figure 5 488 nm**

| **Parameter** | **$V_B^-$ Channel Fit and Output:** | **Het Channel Fit and Output** |
|---|---|---|
| A1 | -49.652 | -138.425 |
| x01 | 3.417e+09 | 3.415e+09 |
| gamma1 | 3.678e+07 | 3.641e+07 |
| A2 | -46.258 | -144.276 |
| x02 | 3.525e+09 | 3.522e+09 |
| gamma2 | 4.245e+07 | 3.502e+07 |
| Offset | 1305.672 | 4074.116 |
| Contrast 1 | 3.80 % | 3.39 % |
| Contrast 2 | 3.54 % | 3.54 % |
| R photons/s | 130600 | 407400 |
| Channel Sensitivity 1 (T/√Hz) | 1.47e-04 | 9.24e-05 |
| Channel Sensitivity 2 (T/√Hz) | 1.82e-04 | 8.51e-05 |
| **Average (T/√Hz)** | **1.65e-04** | **8.85e-05** |

**Table 4 Fit parameters from Figure 5 555 nm**

| **Parameter** | $V_B^-$ **Channel Fit and Output** | **Het Channel Fit and Output** |
|---|---|---|
| A1 | -4.214 | -17.728 |
| x01 | 3.418e+09 | 3.417e+09 |
| gamma1 | 6.455e+07 | 3.778e+07 |
| A2 | -3.574 | -17.304 |
| x02 | 3.534e+09 | 3.521e+09 |
| gamma2 | 2.800e+07 | 4.32e+07 |
| Offset | 351.56 | 825.09 |
| Contrast 1 | 1.20 % | 2.15 % |
| Contrast 2 | 1.02 % | 2.10 % |
| R photons/s | 35100 | 82500 |
| Channel Sensitivity 1 (T/√Hz) | 1.58e-03 | 3.37e-04 |
| Channel Sensitivity 2 (T/√Hz) | 8.08e-04 | 3.95e-04 |
| **Average (T/√Hz)** | **1.19e-03** | **3.66e-04** |

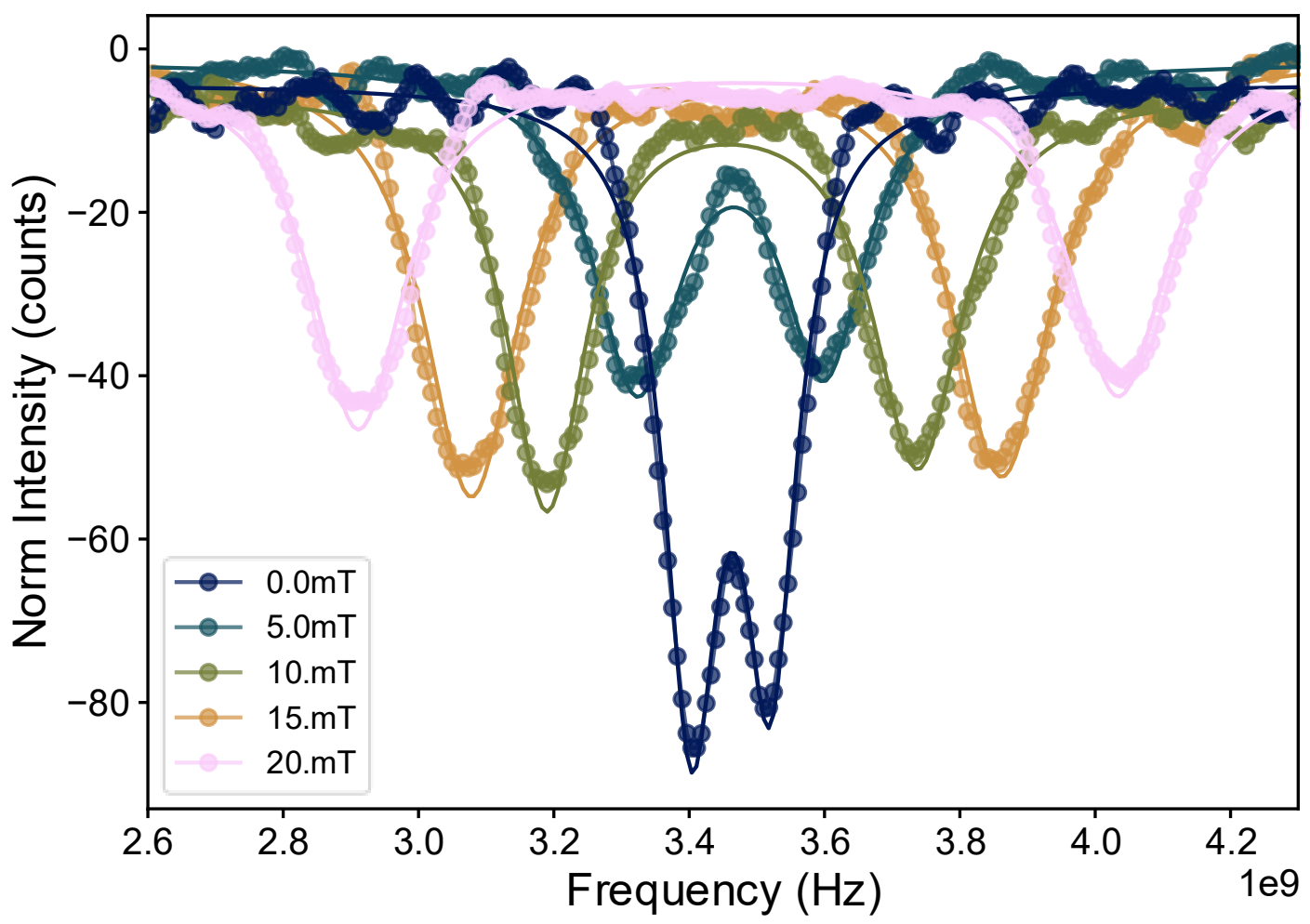

**SI Figure 11 Full ODMR Contrast for Increasing Magnetic Field**: The ODMR spectra prior to 0-1 normalization. This spectrum is computed through an on and off RF pulse scheme. The RF is pulsed on and off at each frequency while the laser stays constant the entire time. The "RF off" counts are subtracted from the "RF on" counts, resulting in the above data. Raw data counts for both on and off are also recorded.

SI Fig. 12 shows that the difference between the sensitivity of the defect in the coupled and isolated structures increases at lower powers. Using the coupled structure, we can achieve better sensitivity with 48 µW excitation power (193 µT Hz$^{-1/2}$) than with 160 **µ**W in the isolated $V_B^-$ (195 µT Hz$^{-1/2}$). After storing in the glovebox for several months, the same heterostructure was used for photostability measurements shown in SI Figure 13, showing the stability of the ODMR sensitivity and enhancement over the course of ten days, while stored outside the glove box between days. This supports the positive effect of capping the structure, discussed in SI Fig. 7.

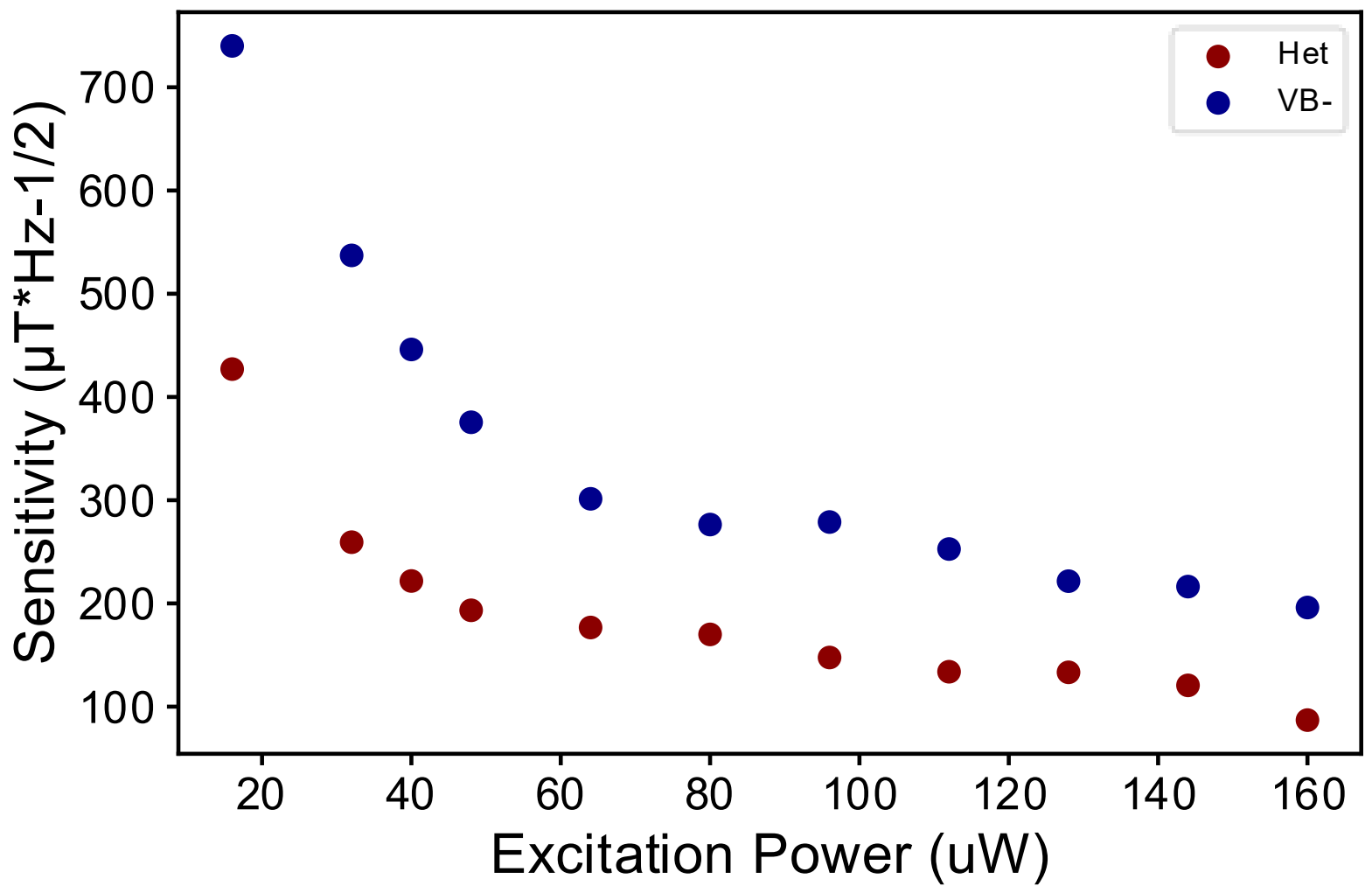

**SI Figure 12 Sensitivity Vs Power**: 488 nm excitation power versus sensitivity (µT Hz$^{-1/2}$) for the heterostructure in Main Text Figure 5 and **SI Figures 3 and 13**.

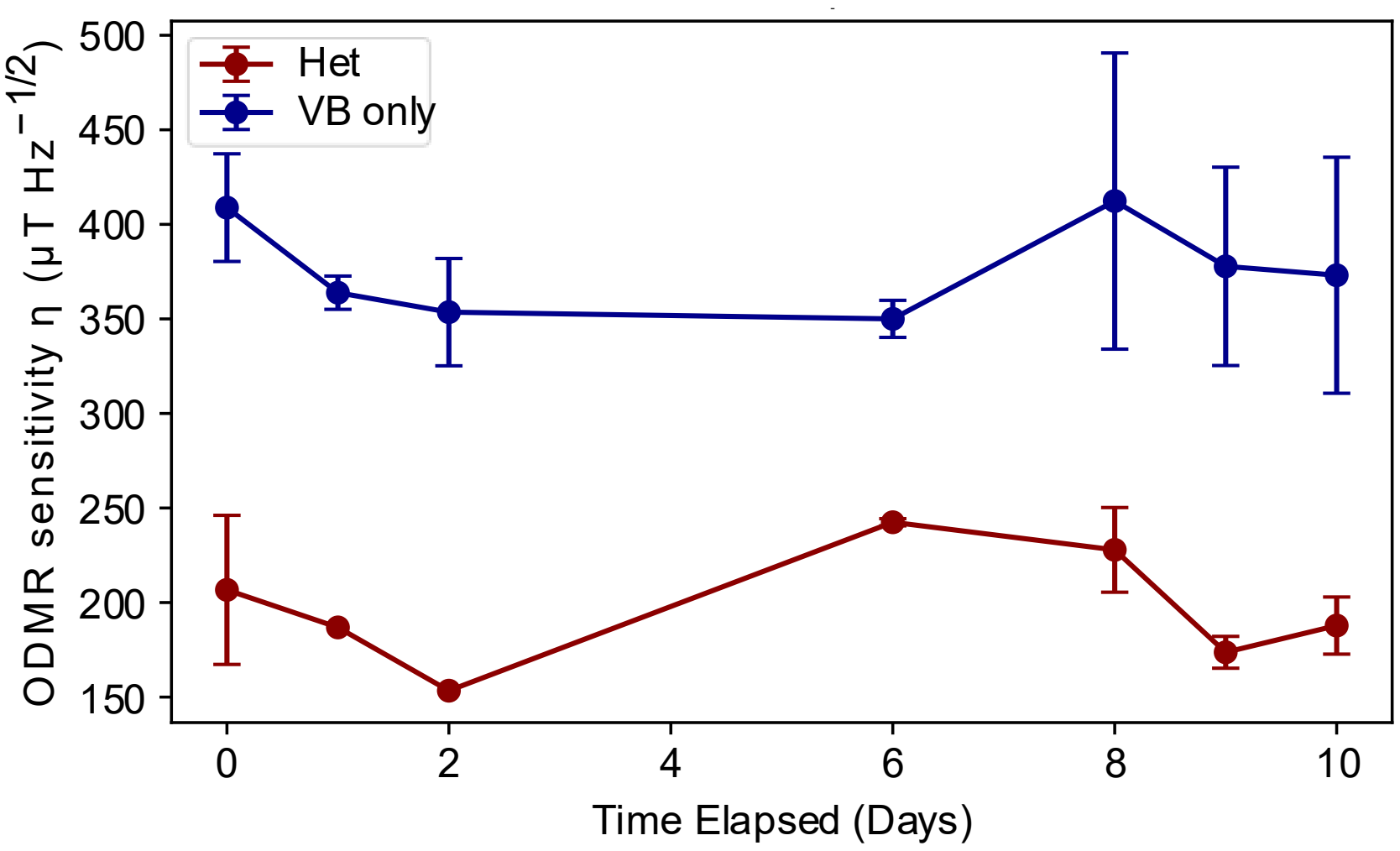

**SI Figure 13 ODMR Sensitivity Stability**: ODMR sensitivity (μT Hz$^{-1/2}$) was measured in each region over the course of 10 days. The standard deviation of replicates from each day is shown by the error bars. The excitation energy was 488 nm with 128 μW power.

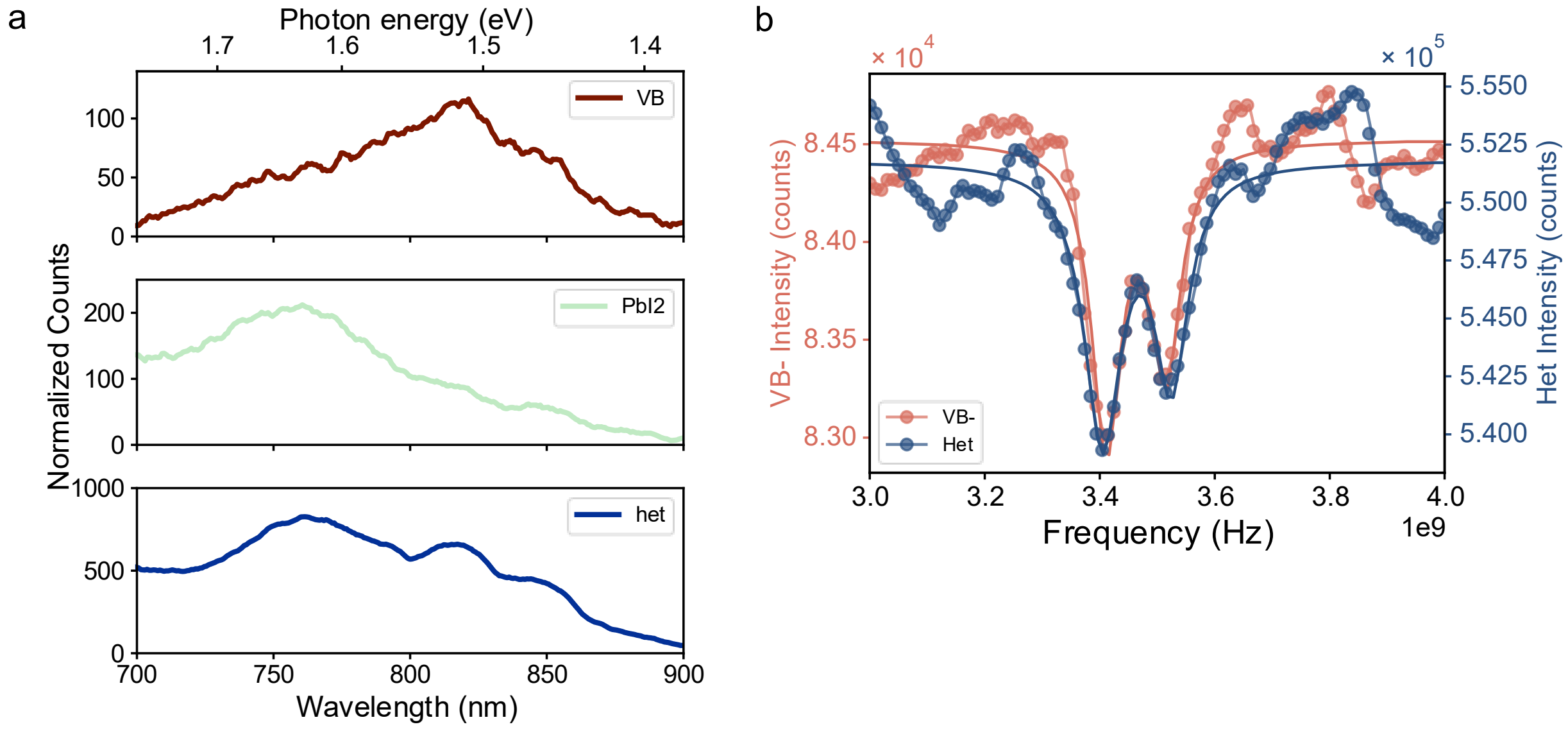

**SI Figure 14 Room Temperature Spectra and ODMR Performance of Device in Main Text Fig. 1**: **a)** Individual room temperature spectra from the heterostructure in Main Text Figure 1, excited with a 488 nm laser. **b)** 488 nm excitation sensitivity (μT Hz$^{-1/2}$) of the heterostructure as compared to the $V_B^-$ region only.

**Table 5 Fit parameters from SI Figure 14 (Device in Main Text Figure 1)**

| **Parameter** | $V_B^-$ **Channel Fit and Output** | **Het Channel Fit and Output** |
|---|---|---|
| A1 | -15.48 | -119.4 |
| x01 | 3.409e+09 | 3.402e+09 |
| gamma1 | 3.056e+07 | 3.836e+07 |
| A2 | -3.574 | -17.304 |
| x02 | 3.521e+09 | 3.522e+09 |
| gamma2 | 2.800e+07 | 4.32e+07 |
| Offset | 845.2 | 5520.7 |
| Contrast 1 | 1.83 % | 2.17 % |
| Contrast 2 | 1.37 % | 1.67 % |
| R photons/s | 84520 | 552100 |
| Channel Sensitivity 1 (T/√Hz) | 3.16e-04 | 1.31e-04 |
| Channel Sensitivity 2 (T/√Hz) | 4.20e-04 | 1.54e-04 |
| **Average (T/√Hz)** | **3.68e-04** | **1.43e-04** |

## 4.2 FROM PL ENHANCEMENT TO ODMR SENSITIVITY

Having identified the origin of the off-overlap emission, we next quantify how the heterostructure affects ODMR sensitivity. The $PbI_2$-associated broadband emission exhibits substantial flake-to-flake variation and spatial heterogeneity, with peak positions ranging from the red to near-infrared. This variation can affect the measured ODMR contrast because red-shifted $PbI_2$ broadband emission may overlap the $V_B^-$ emission window, acting as a spin-independent background that lowers the measured ODMR contrast, thereby degrading the ODMR sensitivity gained from increased photoluminescence intensity.

For sensing, the figure of merit is the ODMR sensitivity rather than the PL intensity. The continuous-wave (CW) ODMR sensitivity scales as $\eta_{\mathrm{CW}} \propto \Delta\nu/(C\sqrt{R})$ (Main Text Eq. 4), with $\Delta\nu$ the resonance linewidth, $C$ the ODMR contrast, and $R$ the detected photon count rate. At fixed linewidth the improvement over bare hBN is $(C/C_0)\sqrt{R/R_0}$, where the subscript 0 denotes the bare-hBN reference. Part of the $PbI_2$ STE emission leaks into the detection window; these photons do not respond to the microwave drive and dilute the contrast. With $E$ the enhancement factor of the $V_{\mathrm{B}}^-$ signal and $\beta = B/R_0$, where $B$ is the STE count rate in the detection window, the total count rate is $R/R_0 = E + \beta$ and the contrast is $C/C_0 = E/(E+\beta)$, so the sensitivity gain is

$$g_{\mathrm{sens}} = \frac{C}{C_0}\sqrt{\frac{R}{R_0}} = \frac{E}{\sqrt{E+\beta}} \,. \tag{9}$$

The factor $E/(E+\beta)$ describes only the STE dilution; changes of the intrinsic contrast are possible in either direction but are not needed for the argument. For $\beta = 0$ the gain reaches the shot-noise ideal $g_{\mathrm{sens}} = \sqrt{E}$; a nonzero STE background caps it below $\sqrt{E}$, and a large $\beta$ erases it even when $E$ is large.

Measured examples of both regimes are analyzed in SI Section 4. Because $\beta$ grows with the pumped $PbI_2$ volume, $\beta \propto 1 - e^{-t_{\mathrm{PbI_2}}/L_{\mathrm{abs}}}$, while $E$ saturates on the few-nanometer coupling scales, the STE overlap rather than the raw photon gain sets the achievable sensitivity. The STE enters this analysis only as a measured quantity ($L_{\mathrm{loss}}$ and $\beta$ are taken from the measured spectra); no microscopic model of self-trapping is required. For sensing applications, the goal is therefore to maximize $E$ while keeping the STE background in the detection window small, via the $PbI_2$ thickness and via spectral filtering.

For example, the structure in SI Fig. 15, shows improved sensitivity (sensitivity specified in the title), but the improvement is smaller (1.4×) than expected (2.8×) from the photon increase alone (8.1×) because, when the $PbI_2$ broadband component enters the $V_B^-$ detection window, it increases the spin-independent background and can dilute the ODMR contrast. This comparison identifies selection criteria for the $PbI_2$ red spectral properties and establishes the suppression of this signal a practical route for converting large photon-count enhancements into correspondingly greater sensitivity improvements.

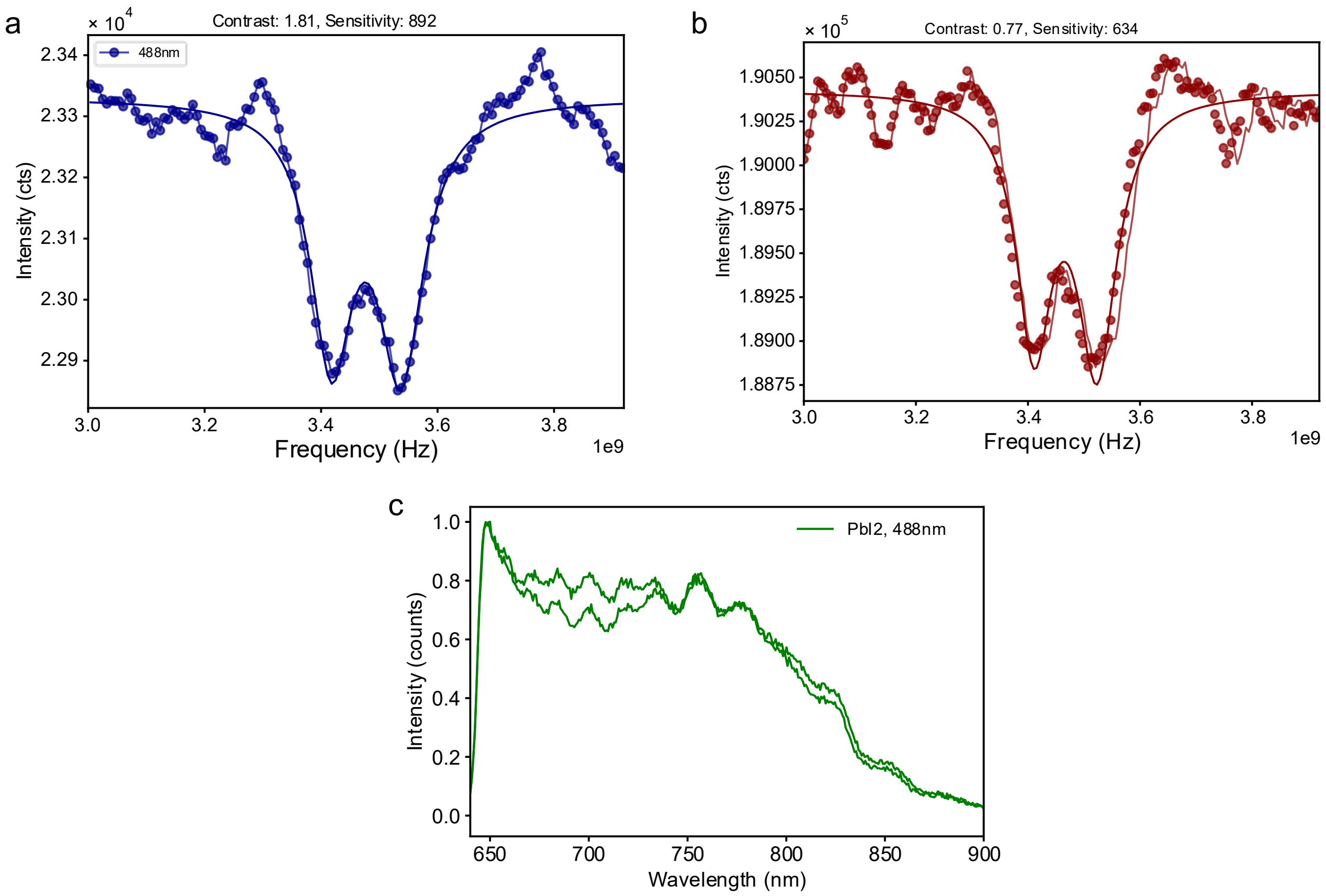

**SI Figure 15 Example Low Contrast Heterostructure due to Redshifted $PbI_2$ Broadband Emission**: **a)** Neutron irradiated region ODMR. Contrast (%) and sensitivity (μT $Hz^{-1/2}$) are in the title. **b)** Heterostructure contrast and sensitivity. **c)** Normalized spectra from $PbI_2$ flake in this heterostructure, showing red-shifted broadband emission from $PbI_2$ which overlaps strongly with $V_B^-$ detection window (775 nm LP filter generally used).

# 5 Discussion on Distance Dependence of Excitation and Energy Transfer

These results identify both spectral background and donor–defect separation as key design parameters; we next consider the expected thickness dependence. The measured thicknesses (SI Fig. 9) do not constitute a systematic thickness series; however, the expected distance dependence can be assessed from the characteristic length scales of the possible coupling channels.

## 5.1 Relevant Length Scales of the Coupling Channels

### 5.1.1 Förster (dipole–dipole) Transfer

For energy transfer between a single $PbI_2$ exciton and a single $V_B^-$ center the efficiency follows the $r^{-6}$ law with the Förster radius

$$R_0^6 = \frac{9(\ln 10)\kappa^2 \Phi_D J}{128\pi^5 n^4 N_A}. \quad (10)$$

The overlap integral, evaluated from the measured spectra (Main Text Figure 3g), is $J = 3.3 \times 10^{12}\ \mathrm{M^{-1}cm^{-1}nm^4}$; the $PbI_2$ emission quantum yield is taken as $\Phi_D = 0.87$, an illustrative near-unity value. For the orientational factor, we take $\kappa^2 = 1/3$, below the isotropic average of $2/3$, since the $PbI_2$ free-exciton dipole is mainly in-plane while the $V_B^-$ transition dipole has a large out-of-plane component. For the refractive index we use $n \approx 2.1$, the in-plane value of hBN in the visible range.[12] This gives $R_0 \approx 1.3$ nm. Since $R_0$ depends on every input only as the $1/6$ power, even large uncertainties (a factor of 5 in $\Phi_D$ together with a factor of 2 in $\kappa^2$) change it by less than a factor of 1.5.

For a $V_B^-$ center at distance $z$ from the $PbI_2$ plane,

$$k_{FRET}(z) \propto \int d^2\rho \, \frac{R_0^6}{(\rho^2 + z^2)^3} = \frac{\pi R_0^6}{2z^4} \tag{11}$$

so the distance dependence softens from $r^{-6}$ to $z^{-4}$. The transfer efficiency becomes

$$\eta_{FRET}(z) = \frac{1}{1 + \left(\frac{z}{z_0}\right)^4}, \qquad z_0 = \left(\frac{\pi}{2}\sigma R_0^6\right)^{\frac{1}{4}} \approx 0.5\text{–}1\ nm \tag{12}$$

for areal defect densities $\sigma = n_{V_\mathrm{B}} d \sim 10^{12}$–$10^{13}$ cm$^{-2}$ per hBN layer, obtained from the interlayer spacing $d = 0.33$ nm and illustrative volumetric vacancy densities $n_{V_\mathrm{B}} \sim 3 \times 10^{19}$–$3 \times 10^{20}$ cm$^{-3}$ for our irradiation dose of $2.6 \times 10^{16}$ n cm$^{-2}$. Since $z_0$ scales only as $\sigma^{1/4}$, even a factor-of-10 uncertainty in the defect density changes it by less than a factor of 1.8. The plane integration also removes most of the sensitivity to the single-pair $\kappa^2$, because laterally displaced pairs contribute with nonzero orientational factors even for orthogonal dipoles.

#### 5.1.2 Dexter (exchange) **T**ransfer

Dexter transfer requires direct wavefunction overlap and decays exponentially, $k_{\mathrm{DET}} \propto \exp(-2r/L)$, with $L$ of the order of the van der Waals contact distance. Across the ≈ 3.4 Å gap, this limits Dexter transfer to the first one or two hBN layers, an effective range below 1 nm.

#### **5.1.3 Exciton Harvesting via the Type-I Alignment**

Excitons generated in the irradiated hBN itself can diffuse and be trapped by $V_B^-$ centers; the relevant scale is $L_D^{\mathrm{hBN}} = \sqrt{D\tau}$, with $\tau$ the lifetime before trapping (distinct from the $PbI_2$ diffusion length $L_D^{\mathrm{PbI_2}}$ used in section 5.1.1. A Drude estimate $D \sim k_B T \tau_s / m^*$ with $\tau_s = 10$–$100$ fs and $m^* \sim m_e$ gives $D \approx 0.5$–$5$ cm$^2$s$^{-1}$, but this is a band-like upper bound; in the heavily irradiated material transport is hopping-like between localized states, and $D$ is orders of magnitude smaller. With illustrative values spanning the expected range, $D = 10^{-4}$–$10^{-2}$ cm$^2$s$^{-1}$ and $\tau = 1$–$50$ ps, $L_D^{\mathrm{hBN}} \approx 0.1$–$7$ nm; even the upper end is well below the wavelength scale.

### 5.1.4 Nonlinear Photon-Mediated Coupling (Photon Ratchet)

This is considered the main mechanism of the manuscript, and unlike the estimates above it is computed *ab initio*. The susceptibility $\chi^{(2)}(-\Omega; \omega_1, -\omega_2)$ underlying Main Text Fig. 2e is evaluated for the heterostructure supercell that contains the defect, so the defect transition enters the matrix elements directly and its resonant excitation at $\Omega = \omega_1 - \omega_2$ is a computed property of the coupled system, not a free parameter. Only the pump at $\omega_1$ is applied in the experiment; $\omega_2$ is the emitted photon, so the defect is excited by a spontaneous difference-frequency process with $\Omega$ at the $V_B^-$ transition. This channel is classified as a second-order, far-field optical response: the emitted intensity $I(\Omega) \propto \left|\mathbf{P}^{(2)}(\Omega)\right|^2$ is the radiation of the driven macroscopic polarization. The thickness dependence, however, is set by the source of this response, not by the radiated wave. Thin-film interference varies with layer thickness on the scale of a quarter optical wavelength in the material, $\lambda/4n \approx 50\text{–}120$ nm—the shortest distance over which an interference factor can change from maximum to minimum—and its amplitude is bounded by factors of order unity. Over the few-nanometer coupling range it is therefore essentially constant and cannot produce a spectrally resonant, order-of-magnitude enhancement. The enhancement instead originates in the source. For the process to be resonant at the defect transition, the perturbation chain in $\chi^{(2)}$ must pass through the defect state: schematically,

$$A \propto \frac{\langle d|p|c_{PbI_2}\rangle \langle c_{PbI_2}|p|v\rangle}{\text{energy denominators}} \tag{13}$$

and the chain contains a matrix element between a defect orbital localized in hBN and a Bloch state localized in $PbI_2$,

$$\langle d|p|c_{PbI_2}\rangle = \int \phi_d^*(\mathbf{r})\, p\, \psi_{PbI_2}(\mathbf{r})\, d^3\mathrm{r}\,. \tag{14}$$

This is an overlap integral: inside the hBN gap both wavefunctions decay exponentially, with decay lengths $\gamma^{-1}$ of a few tenths of a nanometer, so moving the defect one hBN layer deeper shrinks its evanescent tail at the interface by $e^{-\gamma d}$ with $d = 3.3$ Å. The amplitude therefore falls off as $e^{-\gamma z}$ and the excitation rate as $e^{-2\gamma z}$, on the same sub-nanometer scale as the Dexter channel. The photon operators in the process are long-ranged; the electronic matrix element is not. A defect deeper in the hBN still radiates, but at its bare-hBN rate; the ensemble-averaged enhancement therefore decays as in the toy model below, even though every emitted photon is far-field. A second-order response thus does not imply that the enhancement persists to hundreds of nanometers.

The remaining possibility is a two-step process in which the $PbI_2$ layer radiates a difference-frequency wave and a distant defect absorbs it; this is bounded separately: a nonlinear current confined to a plane radiates the relative field amplitude $|E_{\mathrm{rad}}/E_{\mathrm{pump}}| = (\Omega/2c)\,|\chi^{(2)}_{2\mathrm{D}} E_{\mathrm{pump}}|$, which at the 2D upper bound $|\chi^{(2)}_{2\mathrm{D}}| \sim 1\text{–}5\ \mathrm{nm}^2/\mathrm{V}^3$ and our CW intensities of $\lesssim 10^5\ \mathrm{W\,cm}^{-2}$ is $\lesssim 10^{-5}$, negligible. This does not contradict the negligible hybridization of comment 1: hybridization is second order in the weak interlayer coupling, while the $\chi^{(2)}$ matrix elements enter linearly and resonantly, so a coupling too weak to hybridize the defect states can still open an excitation channel.

For the thickness model below, we use the exponential envelope $P^{(2)}(\Omega, z) = P_0^{(2)} e^{-z/\ell_{\mathrm{NL}}}$ with $\ell_{\mathrm{NL}} \approx 2\text{–}5\ \mathrm{nm}$, which over-bounds the sub-nanometer range by an order of magnitude; the emitted intensity $\propto e^{-2z/\ell_{\mathrm{NL}}}$ drops to $1/e$ within $1-2.5$ nm of the interface. Using the sub-nanometer scale instead changes none of the conclusions below.

The AFM-correlated data support this short-range picture (SI Fig. 9): every measured heterostructure, with irradiated-hBN thicknesses of several tens of nanometers, shows a clear PL enhancement, and none shows the wavelength-scale thickness dependence that a far-field mechanism would imply. The devices also differ in defect density, $PbI_2$ flake quality, STE background, and optical collection, so the enhancement is robust across this diversity; extracting a precise decay length would require a dedicated thickness series, a natural direction for future work. The individual coupling channels share the same few-nanometer scale, so the conclusions of the manuscript hold regardless of which channel dominates.

## 5.2 TOY MODEL OF THE THICKNESS DEPENDENCE

We quantify the expected thickness dependence with a minimal model containing the two channels of distinct functional form, plane-integrated FRET and the nonlinear coupling. The type-I alignment enters as the energetic condition enabling transfer, and Dexter transfer (range below 1 nm) is subsumed in the first-monolayer contribution.

A $V_B^-$ center at depth $z$ below the interface couples with efficiency

$$\eta_{coupling}(z) = w_{FRET}\frac{1}{1+\left(\frac{z}{z_0}\right)^4} + w_{NL}e^{-\frac{2z}{\ell_{NL}}}, \qquad w_{FRET} + w_{NL} = 1 \tag{15}$$

Neutron irradiation creates vacancies uniformly over the hBN thickness, so the ensemble average is

$$\langle \eta_{coupling} \rangle_{t_{hBN}} = \frac{1}{t_{hBN}}\int_0^{t_{hBN}} \eta_{coupling}(z)\, dz. \tag{16}$$

The $PbI_2$ layer enters via the fraction of generated excitons that reach the interface. The pump enters from the hBN side, so exciton generation in $PbI_2$ follows a Beer–Lambert profile $G(z') \propto e^{-z'/L_{\mathrm{abs}}}$, with $z'$ measured from the interface into $PbI_2$ and $L_{\mathrm{abs}} = 1/\alpha \approx 100$ nm for the above-gap absorption coefficient $\alpha \sim 10^5\ \mathrm{cm}^{-1}$ of $PbI_2$. Weighting $G(z')$ by the probability $e^{-z'/L_D^{\mathrm{PbI_2}}}$ that an exciton generated at depth $z'$ reaches the interface, integrating over the layer, and normalizing to the thick-layer limit gives

$$A(t_{PbI_2}) = 1 - e^{-\frac{t_{PbI_2}}{L_{eff}}}, \qquad L_{eff} = \left(L_{abs}^{-1} + \left(L_D^{PbI_2}\right)^{-1}\right)^{-1} \approx 9.1\ nm \tag{17}$$

and with thick-layer losses (self-trapped-exciton formation, reabsorption; scale $L_{\mathrm{loss}}$)

$$A_{loss}(t_{PbI_2}) = \left(1 - e^{-\frac{t_{PbI_2}}{L_{eff}}}\right)e^{-\frac{t_{PbI_2}}{L_{loss}}} \tag{18}$$

The nonlinear channel depends on the $PbI_2$ thickness differently: it requires no exciton population, and its source—the mixed interfacial transitions of item 4—involves $PbI_2$ states of the first one or two layers at the interface. Its thickness factor therefore saturates as soon as those layers are present and is insensitive to the excitonic thick-layer losses,

$$A_{NL}(t_{PbI_2}) = 1 - e^{-\frac{t_{PbI_2}}{\ell_P}} \tag{19}$$

with $\ell_P \approx 1$ nm (one to two $PbI_2$ layers); every conclusion below is unchanged for any $\ell_P$ between 0.7 and 5 nm. The total relative enhancement, with each channel weighted by its own $PbI_2$-thickness factor, is

$$\eta_{total}(t_{PbI_2}, t_{hBN}) = \frac{1}{t_{hBN}} \int_0^{t_{hBN}} \left[ w_{FRET} A_{loss}(t_{PbI_2}) \frac{1}{1+\left(\frac{z}{z_0}\right)^4} + w_{NL} A_{NL}(t_{PbI_2}) e^{-\frac{2z}{\ell_{NL}}} \right] dz \tag{20}$$

Parameters: $z_0 = 1$ nm, $\ell_{\mathrm{NL}} = 5$ nm, and $\ell_P = 1$ nm are the conservative, long-range ends of the estimates in 5.1; the $PbI_2$-side lengths $L_{\mathrm{abs}} = 100$ nm, $L_D^{\mathrm{PbI_2}} = 10$ nm, and $L_{\mathrm{loss}} = 150$ nm are illustrative values of the expected order of magnitude. The weights $w_{\mathrm{FRET}} = 0.35$, $w_{\mathrm{NL}} = 0.65$ are likewise illustrative, not fit parameters; both kernels decay on the same few-nanometer scale, so the predicted thickness dependence is insensitive to their ratio.

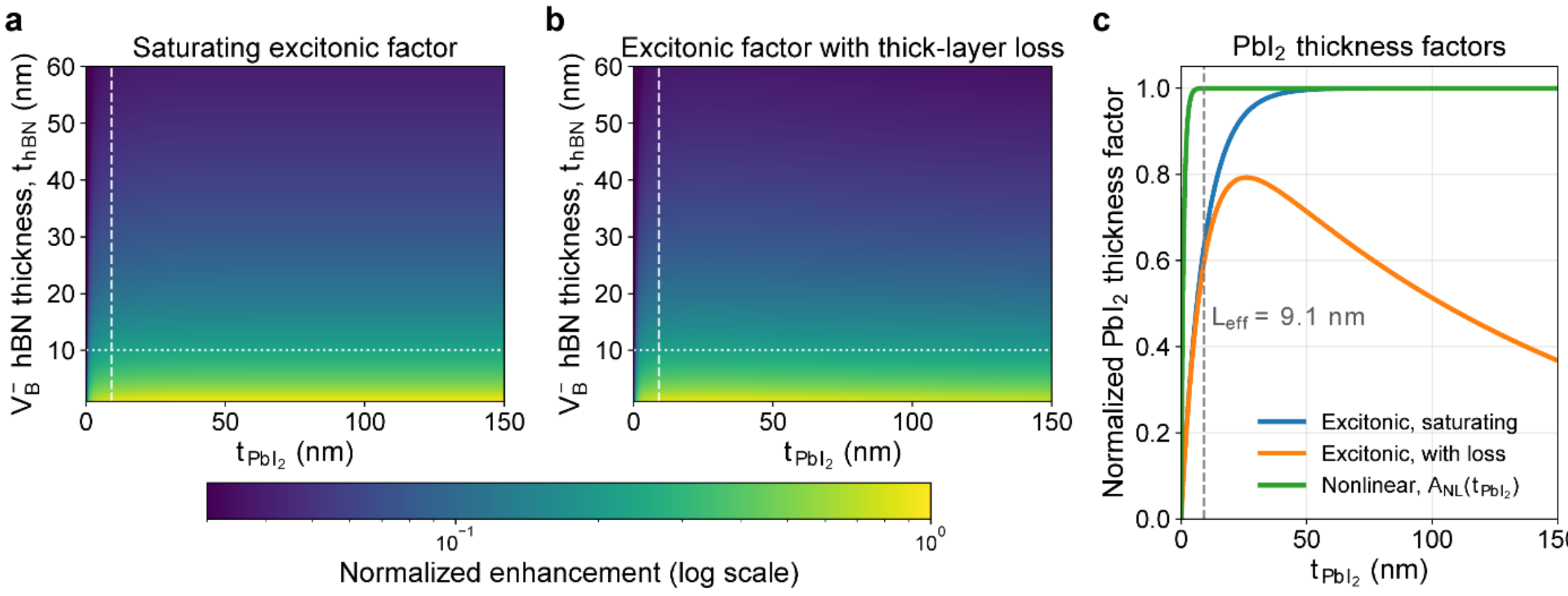

**Figure 16 Toy Model of the Thickness Dependence: a)** Normalized total enhancement $\eta_{\mathrm{total}}(t_{\mathrm{PbI_2}}, t_{\mathrm{hBN}})$ with the saturating excitonic factor $A(t_{\mathrm{PbI_2}})$ and **b)** with the thick-layer loss factor $A_{\mathrm{loss}}(t_{\mathrm{PbI_2}})$; both maps share one color scale, plotted on a log scale. The dashed line marks $L_{\mathrm{eff}} = 9.1$ nm

and the dotted line $t_{\mathrm{hBN}} = 10$ nm. **c)** $PbI_2$-side thickness factors: excitonic $A(t_{\mathrm{PbI_2}})$ without and with losses, and the nonlinear factor $A_{\mathrm{NL}}(t_{\mathrm{PbI_2}}) = 1 - e^{-t_{\mathrm{PbI_2}}/\ell_P}$. Parameters as in the text.

Two predictions follow: the ensemble-averaged enhancement falls off on a few-nanometer hBN scale, and the enhancement persists for thick $PbI_2$ at $\approx 0.8$ of its peak rather than vanishing, because the interfacial nonlinear source and the saturated excitonic factor both survive. These conclusions do not depend on the parameter choices: any nonlinear range between the sub-nanometer overlap scale and the 5 nm envelope, and any $\ell_P$ between $0.7$ and $5$ nm, gives the same behavior. The only parameter-sensitive feature is the position of the shallow $PbI_2$ optimum, which we therefore do not quote. The model results are shown in SI Fig. 16: the enhancement is concentrated at $t_{\mathrm{hBN}} \lesssim 10$ nm for all $PbI_2$ thicknesses [panels (a,b)], and panel (c) shows the saturation of the two $PbI_2$-side factors.

# 6 SIMPLIFIED CALCULATION OF HYBRIDIZATION/SOC EFFECTS DUE TO PROXIMITY OF $PbI_2$

Finally, we assess whether proximity to $PbI_2$ substantially alters the $V_B^-$ spin-state composition. A 7x7 Hamiltonian $\hat{H}$ was constructed for the spin-1 system using 7 basis vectors $\mid j\rangle$ : (3 for each ground state spin projection, 3 for each excited state spin projection, and 1 for the singlet state |g,+1⟩ |g,0⟩ |g,-1⟩ |e,+1⟩ |e,0⟩ |e,-1⟩ |S=0⟩.

$$\hat{H} = \begin{bmatrix} \frac{D_g}{3}+E_g & 0 & E_g & 0 & 0 & 0 & 0 \\ 0 & \frac{-2D_g}{3} & 0 & 0 & 0 & 0 & 0 \\ E_g & 0 & \frac{D_g}{3}-E_g & 0 & 0 & 0 & 0 \\ 0 & 0 & 0 & \frac{D_e}{3}+E_e & 0 & E_e & \lambda_{intr}^* \\ 0 & 0 & 0 & 0 & \frac{-2D_e}{3} & 0 & 0 \\ 0 & 0 & 0 & E_e & 0 & \frac{D_e}{3}-E_e & \lambda_{intr}^* \\ 0 & 0 & 0 & \lambda_{intr} & 0 & \lambda_{intr} & E_s \end{bmatrix}$$

Each eigenvector, $\mid \Psi_i\rangle$ , is a mixture of all 7 basis states, where $C_{i,j}$ are the normalized mixing coefficients (squares sum to 1). Explicitly, $C_{ij}^2$ is the probability of the eigenvector $\mid \Psi_i\rangle$ containing basis state $\mid j\rangle$ .

$$\hat{H} \mid \Psi_i\rangle = E_i \mid \Psi_i\rangle, i = 0, \dots, 6,$$

$$|\Psi_i\rangle = \sum_{j=0}^{6} c_{ji} \mid j\rangle$$

Eigenvalues and eigenvectors were obtained by diagonalizing $\hat{H}$. The coefficients for each vector were extracted. This was done with and without the presence of nearby lead iodide, using an intrinsic SOC ~ 170 GHz.[13]

The presence of heavy atoms such as Pb and I gives rise to relativistic spin–orbit coupling (SOC) that modifies the band structure and carrier dynamics compared to non-SOC calculations. First-principles studies show that including SOC from heavy elements alters orbital mixing and band energies by hundreds of meV and suppresses recombination rates,[14,15] consistent with effective SOC enhancements on the order of 20–50% used in this approximation. These effects originate from wavefunction hybridization between defect states and Pb/I orbitals, which increases the orbital angular momentum character and introduces proximity-induced relativistic contributions into the defect's effective SOC. As a first approximation in our simplified model, we adopt an isotropic/symmetrical enhancement factor to scale SOC matrix elements, thereby estimating the maximum impact of heavy-atom proximity on spin–orbit mediated mixing. In reality, this enhancement would not be isotropic, and therefore we would like to emphasize that this is a qualitative investigation.

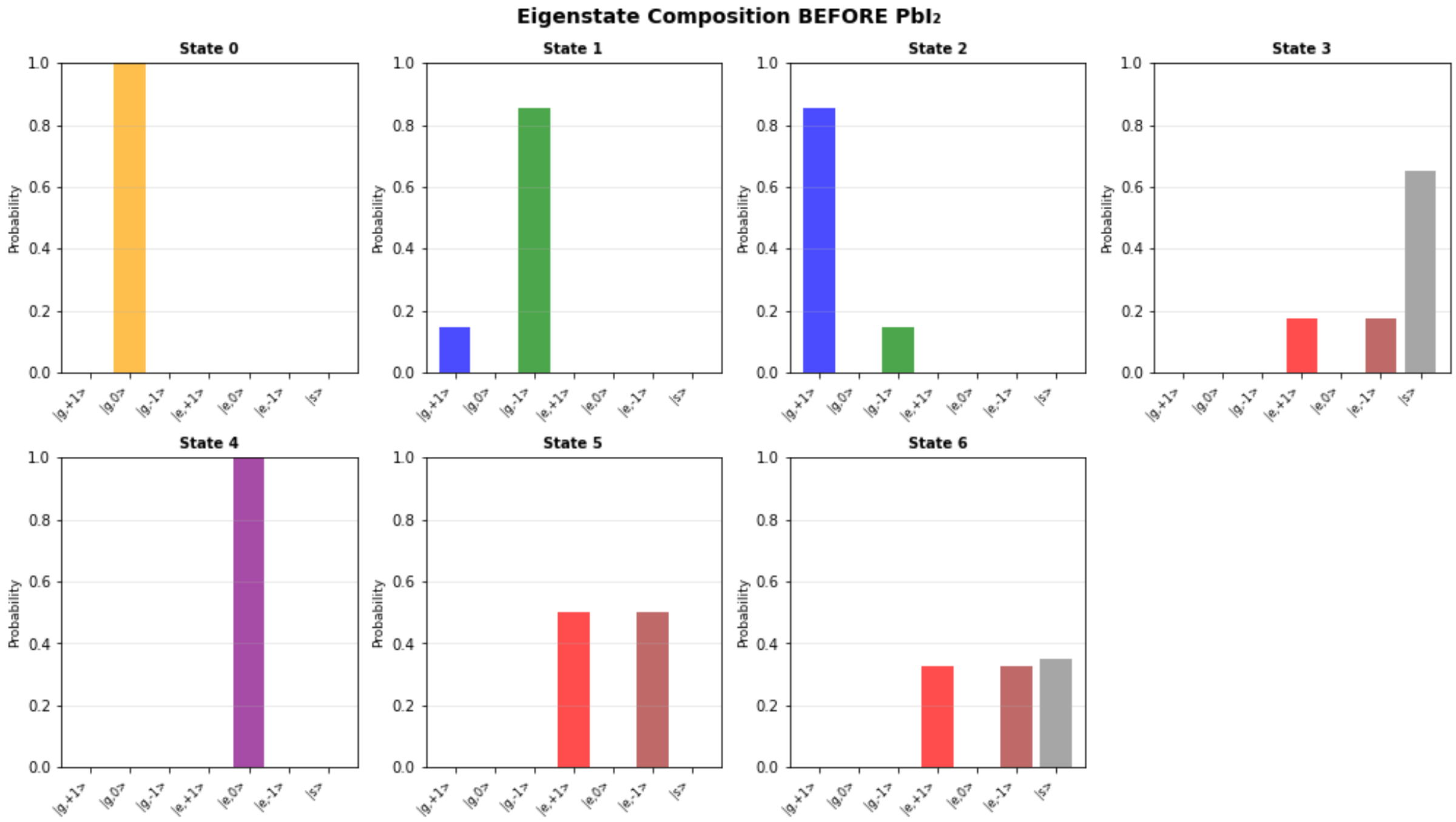

**SI Figure 17 Eigenvectors and Basis State Probabilities Before $PbI_2$ Coupling**

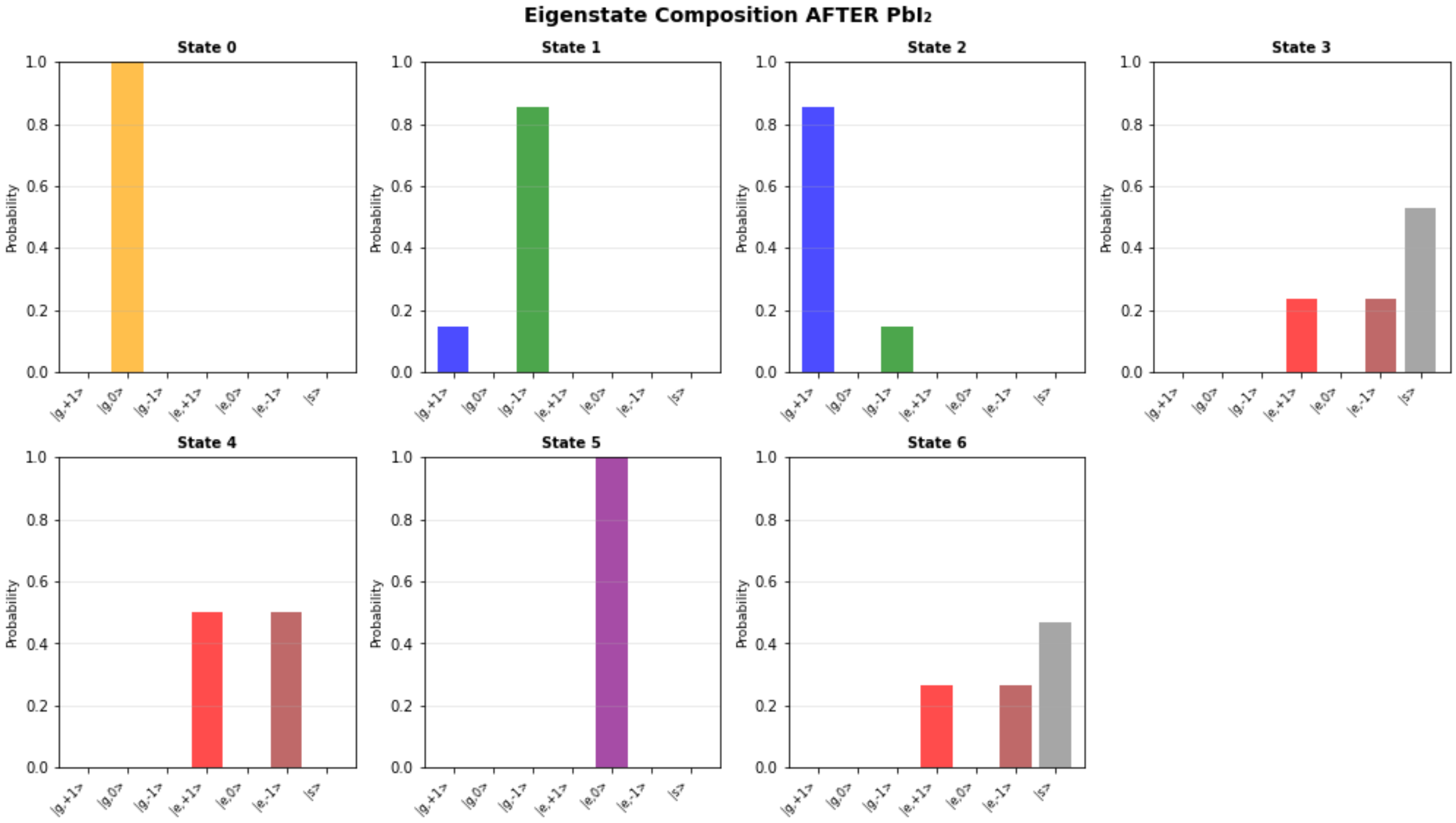

**SI Figure 18 Eigenvectors and Basis State Probabilities After $PbI_2$ Coupling**

As can be observed, the states are reorganized, but the composition probabilities remain largely unchanged. Since there is little difference between the probability that each eigenvector acts as each basis state, this predicts, coarsely, that little hybridization takes place. Diagonalizing $\widehat{H}$ mixes the pure spin states into "new states" (eigenvectors). Each new state has different weights $\left(C_{ij}^{2}\right)$ of the original pure states, but $PbI_2$ barely changes these mixing amounts, supporting the retention of spin-selective rules required to have ODMR.

# 7 Bibliography of SI